\documentclass[%
 reprint,
 amsmath,amssymb,
 aps,
prb,
longbibliography,
]{revtex4-1}

\usepackage{graphicx}
\usepackage{dcolumn}
\usepackage{bm}
\usepackage{float} 
\usepackage{lipsum}
\usepackage{enumitem,amssymb} 
\usepackage{color}
\usepackage[usenames,dvipsnames,svgnames,table]{xcolor}
\usepackage{hyperref}
\hypersetup{
    colorlinks = true,
    citecolor = blue
}
\newlist{todolist}{itemize}{2}
\setlist[todolist]{label=$\square$}
\usepackage{pifont}

\usepackage{leftidx}
\usepackage{csquotes}
\usepackage{tabulary}
\usepackage{tabularx,booktabs}
\usepackage{subcaption} 
\begin{document} 


\preprint{APS/123-QED}
  
\title{Influence of finite temperature Exchange-Correlation effects in Hydrogen }
\author{Kushal Ramakrishna\textsuperscript{1,2,}}
\email{k.ramakrishna@hzdr.de}
\author{Tobias Dornheim\textsuperscript{3}}
\author{Jan Vorberger\textsuperscript{1}}
\affiliation{\textsuperscript{1}Helmholtz-Zentrum Dresden-Rossendorf, Bautzner Landstra\ss e 400, 01328 Dresden, Germany}
\affiliation{\textsuperscript{2}Technische Universit\"at Dresden, 01062 Dresden, Germany}
\affiliation{\textsuperscript{3}Center for Advanced Systems Understanding (CASUS), G\"orlitz, Germany}

\date{\today}

\begin{abstract}  
We use density functional molecular dynamics (DFT-MD) to study the effect of finite temperature exchange-correlation (xc) in hydrogen. Using the Kohn-Sham approach, the xc energy of the system, $E_\textnormal{xc}(r_{s})$ is replaced by the xc free energy $f_\textnormal{xc}(r_s,\Theta)$ within the local density approximation (LDA) based on parametrized path integral Monte Carlo (PIMC) data for the uniform electron gas (UEG) at warm dense matter (WDM) conditions. We observe insignificant changes in the equation of state (EOS) at the region of metal-insulator transition compared to the regular LDA form, whereas significant changes are observed for T$>$10000 K, i.e., in the important WDM regime. 
Thus, our results further corroborate the need for temperature-dependent xc functionals for DFT simulations of WDM systems. Moreover, we present the first finite-temperature DFT results for the EOS of hydrogen in the electron liquid regime up to $r_{s}=14$ and find a drastic impact (about $20\%$) of thermal xc effects, which manifests at lower temperatures compared to WDM. 
We expect our results to be important for many applications beyond DFT, like quantum hydrodynamics and astrophysical models.
\end{abstract}  

\pacs{Valid PACS appear here}
\maketitle 

\section{Introduction}  
 
Over the last decade or so, the interest on the properties of matter under extreme conditions has drastically increased due to new experimental and theoretical methods~\cite{Fortov_2009}. Of particular importance is the so-called warm dense matter (WDM) regime~\cite{dornheim_review,2019arXiv191209884B}, which is defined by two characteristic parameters being of order unity: 1) the density parameter $r_s=\overline{r}/a_\textnormal{B}$ (with $\overline{r}$ and $a_\textnormal{B}$ being the average inter-particle distance and first Bohr radius, respectively) and 2) the reduced temperature $\Theta=k_\textnormal{B}T/E_\textnormal{F}$ (with $E_\textnormal{F}$ being the Fermi energy~\cite{giuliani_vignale_2005}). We note that the latter can be viewed as a degeneracy parameter~\cite{Ott2018}, and $\Theta\ll1$ ($\Theta\gg1$) indicates when a $T=0$ description
(a non-degenerate treatment) of the electronic subsystem is appropriate.    


In nature, WDM occurs in astrophysical objects such as brown and white dwarfs~\cite{saumon_the_role_1992,chabrier_quantum_1993,chabrier_cooling_2000}, interiors of giant planets~\cite{nettelmann_saturn_2013,militzer_massive_2008,vorberger_hydrogen-helium_2007}, and meteor impacts~\cite{MELOSH1984234,dlott}. Moreover, these conditions can be realized experimentally using different methods such as laser or ion beam compression~\cite{glenzer_matter_2016,tschentscher_photon_2017,hoffmann_cpp_18}, see Ref.~\onlinecite{falk_2018} for a topical review article. Finally, we mention that WDM is predicted to occur on the pathway towards inertial confinement fusion~\cite{PhysRevB.84.224109,matzen_pulsed-power-driven_2005}, which makes a thorough understanding of this regime highly desirable.

From a theoretical perspective, the condition $r_s\sim\Theta\sim1$ implies an intricate interplay of quantum scattering, electronic degeneracy effects and thermal excitations, which renders the
accurate description of WDM a formidable challenge. More specifically, there are no small parameters and, hence, perturbative methods break down~\cite{graziani2014frontiers}. This leaves computationally expensive \textit{ab initio} simulations as the only option. In this work, we focus on density functional theory (DFT), which has emerged as the de-facto work horse in modern many-body theory~\cite{Burke_JCP,Jones_RMP}.

In particular, the 
success of DFT regarding the description of real materials was facilitated by the availability of accurate  exchange-correlation (xc) functionals, which, however, cannot be obtained within DFT itself and have to be supplied as input. While the exact functional is, in general, not known, this quantity can often be reasonably approximated on the basis of the properties of the uniform electron gas~\cite{loos_review,dornheim_review} (UEG). In this context, the key quantity is given by the xc-energy of the UEG, which, at zero temperature, was accurately computed by Ceperley and Alder~\cite{ceperley_alder}. These data were subsequently used as input for different parametrizations\cite{vwn,perdew_zunger,perdew_wang} of $E_\textnormal{xc}(r_s)$, which allow for DFT calculations on the level of the local density approximation (LDA). Moreover, these results constitute the basis for more advanced functionals like the celebrated work by Perdew, Burke, and Ernzerhof\cite{pbe} (PBE).

While the generalization of DFT from the ground-state to finite temperature was introduced over 50 years ago by Mermin~\cite{mermin}, most results for WDM  up to this date have been obtained on the basis of the \textit{zero-temperature approximation} (e.g., Refs.~\onlinecite{PhysRevE.71.016409,PhysRevB.81.054103,schottler_redmer_2018,PhysRevB.95.144105,PhysRevLett.116.115004,doi:10.1002/ctpp.201400101}), i.e., using xc-functionals that were designed for the ground-state. However, this assumption is highly questionable, as the thermal DFT formalism requires as input a parametrization of the xc-free energy $f_\textnormal{xc}(r_s,\Theta)$ that explicitly depends both on density and temperature~\cite{gupta_dft,burke_chapter}. Therefore, 
replicating the success of ground state DFT at elevated temperatures requires an accurate description of the UEG in the WDM regime.

This need has sparked a surge of new developments regarding quantum Monte Carlo simulations of electrons in this regime~\cite{Brown_PRL,blunt_prb,Dornheim_2015,Malone_JCP,Dornheim_JCP,Groth_PRB,Dornheim_PRB,Malone_PRL,Dornheim_PRL,clark_prb,Dornheim_POP,PhysRevLett.119.135001,dornheim2019static}. More specifically, Brown \textit{et al.}~\cite{Brown_PRL} presented the first path integral Monte Carlo (PIMC) results for the warm dense UEG, which were subsequently used as input for several parametrizations~\cite{PhysRevB.88.081102,PhysRevB.88.115123,ksdt}. While being an important first step, the Brown \textit{et al.}~\cite{Brown_PRL} data were obtained by imposing a restriction on the nodal structure of the thermal density matrix (\textit{fixed node approximation}~\cite{Ceperley1991}) so that the quality of these data had remained unclear. Shortly thereafter, Schoof \textit{et al.}~\cite{PhysRevLett.115.130402} were able to unambiguously quantify the nodal errors by presenting an independent data set using the exact configuration PIMC method~\cite{doi:10.1002/ctpp.201100012}. 
It was found that the restricted PIMC data exhibit systematic deviations of up to $10\%$.

A first extensive set of unbiased PIMC data was then presented~\cite{Dornheim_PRL} and used to construct a new parametrization of $f_\textnormal{xc}$ (hereafter denoted as GDSMFB~\cite{PhysRevLett.119.135001,misc}), which is employed throughout this work.  For completeness, we mention that the earlier parametrization by Karasiev \textit{et al.}~\cite{ksdt} (KSDT) (and also an improved version thereof presented in Ref.~\onlinecite{PhysRevLett.120.076401}, "corrKSDT") exhibits a comparable accuracy in the relevant WDM regime, see Ref.~\onlinecite{PhysRevB.99.195134} for a recent analysis and Ref.~\onlinecite{dornheim_review} for an extensive review article.

In fact, the KSDT functional was used for an investigation of gradient corrections by Sjostrom and Daligault~\cite{PhysRevB.90.155109} who found that finite-$T$ xc effects start to matter at around $T=10^4$K.
The first thorough investigation of finite-$T$ xc effects was presented in Ref.~\onlinecite{PhysRevE.93.063207}, where the limits of the zero-temperature approximation were pointed out for a few different materials and quantities. Shortly thereafter, the same authors presented a finite-$T$ GGA functional~\cite{PhysRevLett.120.076401} and reported a significant improvement in the principal Hugoniot of shocked deuterium~\cite{PhysRevB.99.214110}.

Yet, an overarching study of the impact of exchange--correlation effects considering different relevant physical regimes (e.g., WDM, electron liquid, etc) is still lacking.

In this work, we aim to fill this gap by carrying out extensive thermal DFT calculations of hydrogen and comparing different zero-temperature approximations to the GDSMFB functional~\cite{PhysRevLett.119.135001}. 
In this context, we mention that hydrogen constitutes the most abundant element in our universe and offers a plethora of interesting physical effects~\cite{RevModPhys.84.1607} such as the liquid-liquid insulator-to-metal phase transition, the \textit{holy grail} of extreme-conditions research~\cite{wigner1935possibility,PhysRevLett.21.1748,PhysRevLett.76.1860,dalladay2016evidence,Knudson1455,Celliers677,dias2017observation,eremets2011conductive}. Further actively investigated questions regarding hydrogen at extreme conditions include ionization potential depression~\cite{doi:10.1063/1.4940313,refId0} and proton crystallization~\cite{PhysRevLett.95.235006,doi:10.1002/ctpp.201100085}.
 
Our new results allow us to unambiguously study the impact of the finite-temperature contributions to the xc functional on various quantities such as the EOS and the electronic density of states over a broad range of temperatures and densities ranging from $r_s=0.8137$ up to $r_s=14$ thus exploring finite-T xc effects in systems in which the electron properties are influenced by  a large range of different coupling strengths all the way to the electron liquid~\cite{2019arXiv191107598D}.  

The paper is organized as follows: 
In Sec II., we discuss the computational methods including the choice of the basis set, system size and the employed \emph{k}-point sampling.
The subsequent Sec III. is devoted to our simulation results, including the metal-insulator transition (Sec. III.A) and the warm dense matter regime (Sec. III.B). In Sec. III.C, we extend the considerations to the previously unexplored electron liquid regime at finite temperature. In addition, we present simulation results for the electronic density of states (DOS) and the density in coordinate space in Secs. III.D and III.E, respectively.
The paper is concluded by a brief summary and outlook in Sec. IV.

\section{Computational details}
\label{comp_methods} 

\begin{figure}[th]  
\begin{center}  
\includegraphics[width=1.0\columnwidth]{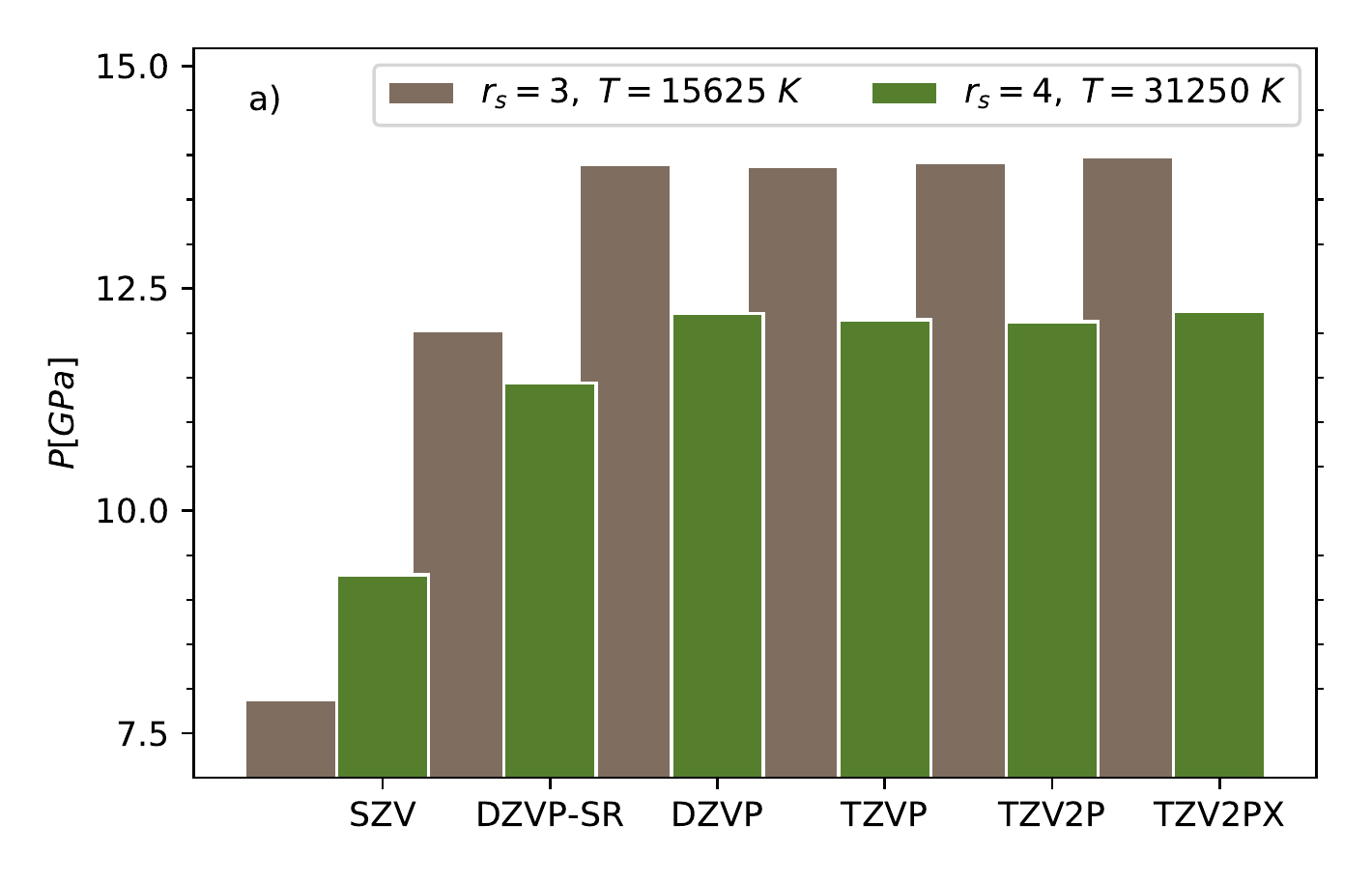} 
\includegraphics[width=1.0\columnwidth]{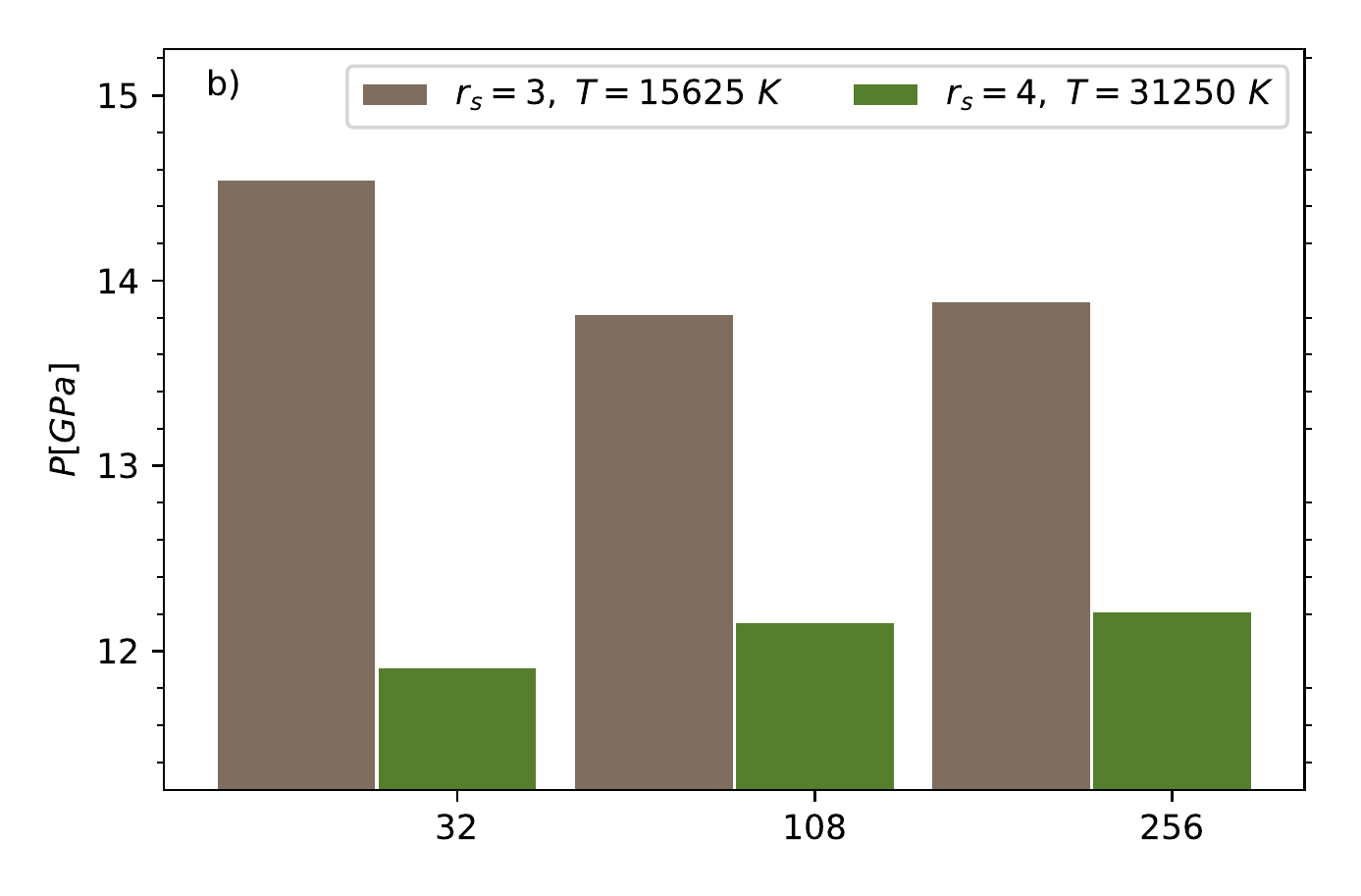} 
\includegraphics[width=1.0\columnwidth]{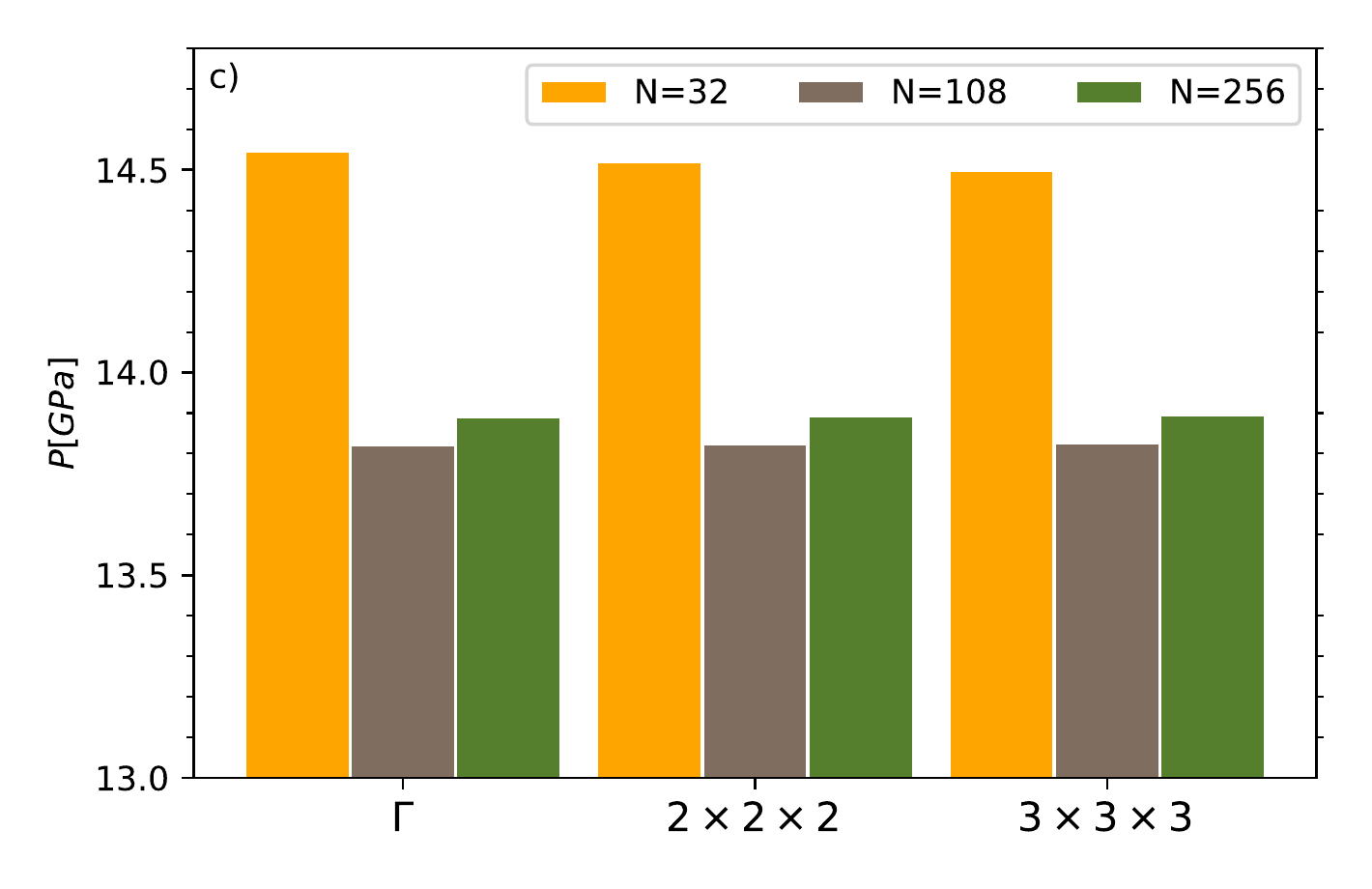}
\end{center}
\caption{\raggedright Pressure variation with respect to a) the basis set at  $r_{s}=3.0$, $T=15625\ K$. The error bars are too small to be shown; b) system size for the DZVP basis set at $r_{s}=3.0$, $T=15625\ K$  and $r_{s}=4.0$, $T=31250\ K$; c) \emph{k}-point sampling for the DZVP basis set at $r_{s}=3.0$, $T=15625\ K$ for a system size of 32, 108 
and 256. 
}  
\label{H2_basisset_vs_pressure}        
\end{figure}

The simulations are performed using the CP2K code~\cite{PhysRevLett.119.135001, doi:10.1002/wcms.1159}. The Kohn-Sham equations are solved using the Gaussian plane waves method with the basis set consisting of Gaussians along with additional plane waves as auxiliary basis. The electron-proton potential is approximated using Goedecker-Teter-Hutter pseudopotentials of LDA form with a cutoff radius of $r_c=0.2\,$a$_B$~\cite{PhysRevB.54.1703}. The standard T$=$0 LDA xc functional is hereafter referred to as PZ~\cite{perdew_zunger} (Perdew-Zunger) and the parametrized temperature dependent LDA form of the xc functional is referred to as GDSMFB~\cite{PhysRevLett.119.135001}. The temperature dependent GDSMFB functional is accessed in the CP2K code using the library of exchange-correlation functionals (LIBXC)~\cite{LEHTOLA20181,MARQUES20122272}.  


\par  
The convergence tests are performed using full DFT-MD simulations by varying the system parameters: \textit{basis set, plane wave energy cutoff, k-point sampling, and system size}. Simulations are run up to 10000 steps and the equilibrated snapshots are averaged to obtain the statistics (pressure/energy) of the system. 
\par  
The choice of the Gaussian basis set is important for obtaining accurate results, while staying within reasonable limits for the computational demand.
The best accuracy with sufficient speed can be obtained when using
the double zeta valence polarized (DZVP) basis set for the computation
of the pressure as summarized in Fig. \ref{H2_basisset_vs_pressure}a for the PZ functional. The choice of the computationally more expensive basis sets can be ignored as the results are converged to within $0.3 \%$. The DZV/DZVP basis sets have been previously utilized in the simulations of warm dense hydrogen and hydrogen-helium mixtures~\cite{doi:10.1063/1.4983057,liu2018quantum}. 
The plane wave energy cutoff of the system is set between 450-800 Ry and the Gaussian basis set cutoff is set to 90-180 Ry depending on the density and the temperature of the system. The convergence with respect to the plane wave energy cutoff and the basis sets are demonstrated in Appendix~\ref{app_low_rs} and Appendix~\ref{app_high_rs} for the highest and lowest density respectively considered in this work. 

Based on the DZVP basis set, the choice of the system size is tested next.
The smallest system size considered for our DFT-MD simulations (N$=32$) shows finite size effects at a range of densities and temperatures and the minimum size required for sufficient accuracy is given by N$=256$ as shown in Figure \ref{H2_basisset_vs_pressure}b. Yet, we are forced to resort to N$=32$ for extreme cases of small/large densities due to high computational demands required, which is explicitly mentioned in those cases. Note that a similar effect of the system size on the EOS is also observed by Lorenzen \textit{et al.} for dense hydrogen~\cite{PhysRevB.82.195107}. The effect is seen in Fig. \ref{H2_basisset_vs_pressure}b for $r_{s}=3.0$ with a pressure variation of $4.7 \%$ with the change in system size from N$=$32 to N$=$256. At $r_{s}=4.0$, the pressure variation with the same change in system size is $2.5 \%$. At even lower densities a system size of N$=32$ or $N=108$ only is feasible for simulations considering the requirement of more plane waves for bigger simulation boxes. 
\par 
The choice of \emph{k}-point sampling can influence the computed energy and pressure of the system. At $r_{s}=3$, $T=15625\ K$, we see no dependence on \emph{k}-point sampling based on the system size for obtaining the convergence as shown in Fig. \ref{H2_basisset_vs_pressure}c. With a small system size of N$=32$, the pressure difference between $\Gamma$-point sampling and the usage of \emph{k}-points are negligible but the pressures obtained are comparatively higher than the values reported by Hu \textit{et al.} and Wang \textit{et al.} respectively~\cite{PhysRevB.84.224109,doi:10.1063/1.4821839}.  For a system size of N$=108$, the pressure difference between $\Gamma$-point sampling and $3 \times 3 \times 3 $ is $<0.1\%$. Higher \emph{k}-point sampling has a smaller effect for bigger supercells as the case of N$=256$ particles demonstrates where the change in pressure is still $<0.1\%$ going from $\Gamma$-point sampling to a $3 \times 3 \times 3 $ grid of \emph{k}-points. For these reasons, we consider a system size of N$=256$ sampled at the $\Gamma$-point.   

  
\par     
 The simulation time step was chosen so as to account for the large kinetic energies of the protons at higher temperatures. It ranges in between 0.02 fs for the highest temperatures and 0.1 fs for the lowest temperatures considered. Simulations ran for at least 10000 steps until the system had equilibrated and then further time steps of 4000-5000 are considered for obtaining the statistics. We use a Fermi occupation of the bands/eigenvalues to set the electronic temperature~\cite{mermin,gupta_dft} and employ a Nos{\'e}-Hoover thermostat to control the ionic temperature in the canonical ensemble~\cite{nose1984unified,nose1984molecular}. The simulation box consisted of a hexagonal cell ($a$=$b$, $c$=1.63$a$) under periodic boundary conditions and the cell size varied depending on the density. The simulations cover the density range from $r_s={0.8137 \ldots 14}$ for a wide range of temperatures $T=250-400000$ K. For completeness, we mention that simulations of temperature ranges beyond 400000 K are at present computationally too expensive using KS-DFT. Alternatives include orbital-free DFT~\cite{PhysRevB.88.161108,KARASIEV20143240,zhang2016extended,PhysRevB.101.075116} and an extended KS-formalism~\cite{zhang2016extended}, which, however, are beyond the scope of this work.  



\section{Results}  

\subsection{ Finite temperature exchange correlation effects in the high pressure fluid }    

\begin{figure}[th]
\centering
\includegraphics[width=1.0\columnwidth]{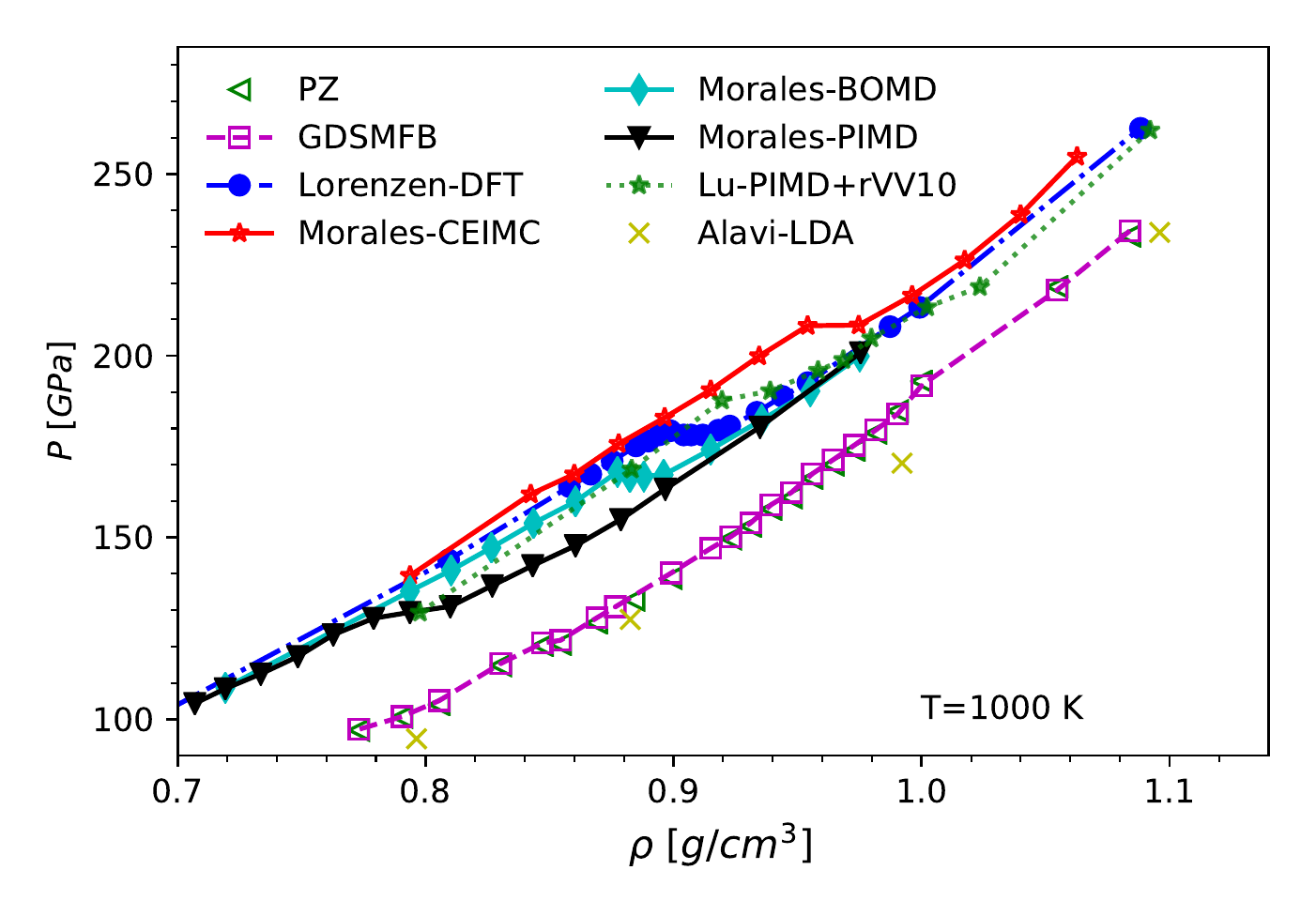}      
\caption{\raggedright EOS at $T=1000 K$ using various xc functionals and \textit{ab initio} methods. Lorenzen Ref.~\onlinecite{PhysRevB.82.195107}; Morales Ref.~\onlinecite{Morales12799}; Lu Ref.~\onlinecite{Lu_2019}; Alavi Ref.~\onlinecite{Alavi1252}. PZ and GDSMFB results of this work. }     
\label{H2_1000}        
\end{figure}   

\begin{figure}[th] 
\centering 
\includegraphics[width=1.0\columnwidth]{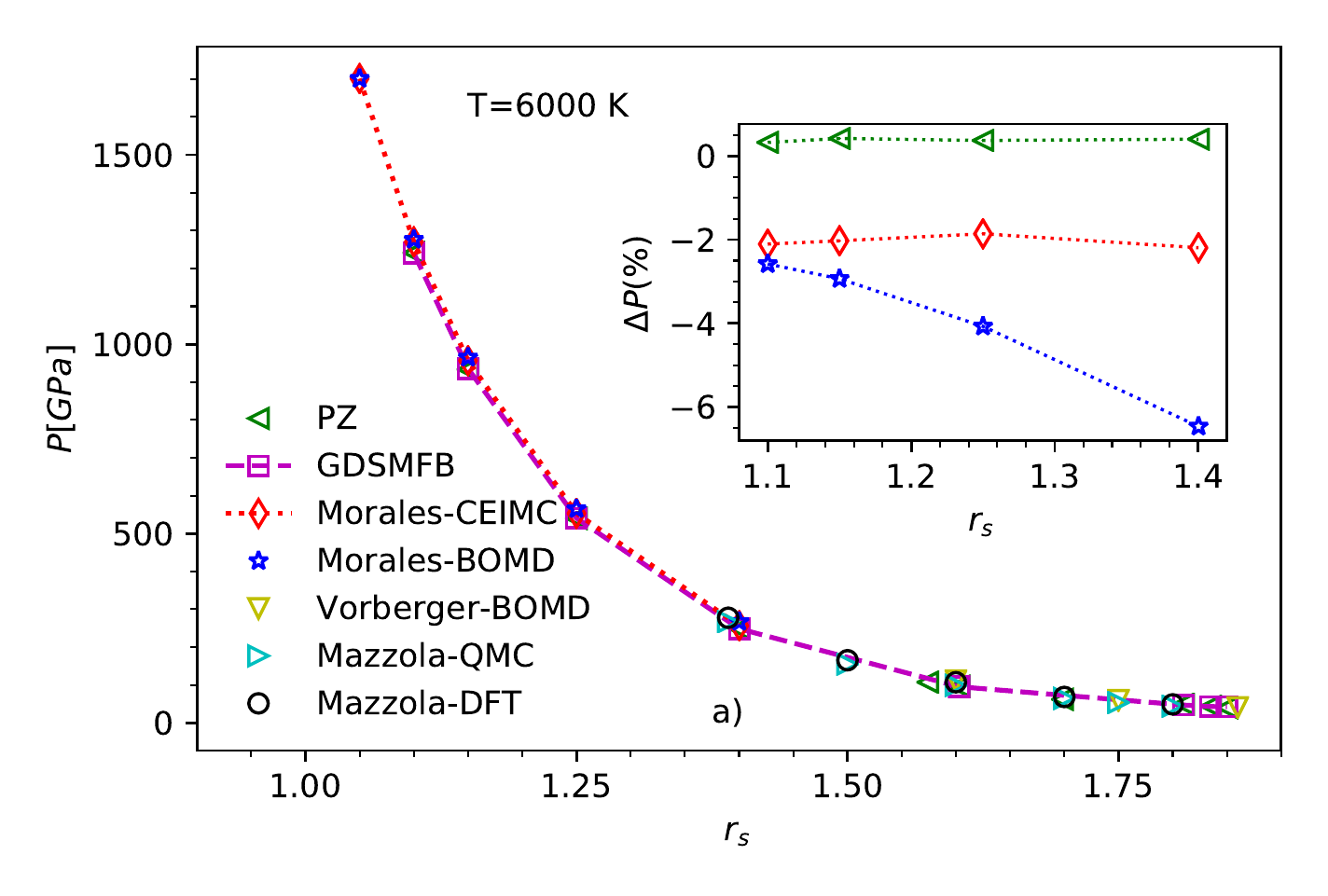}
\includegraphics[width=1.0\columnwidth]{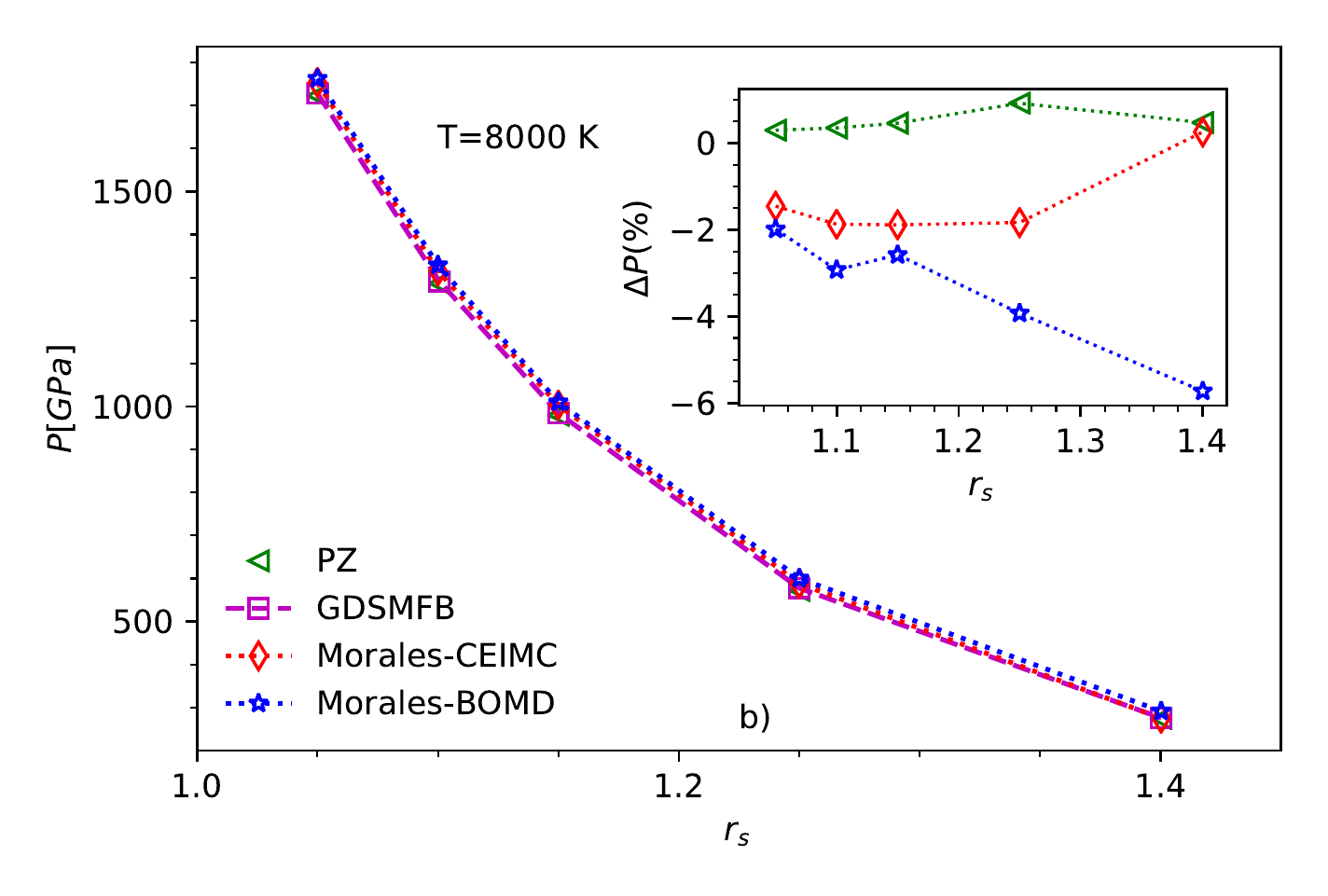}
\includegraphics[width=1.0\columnwidth]{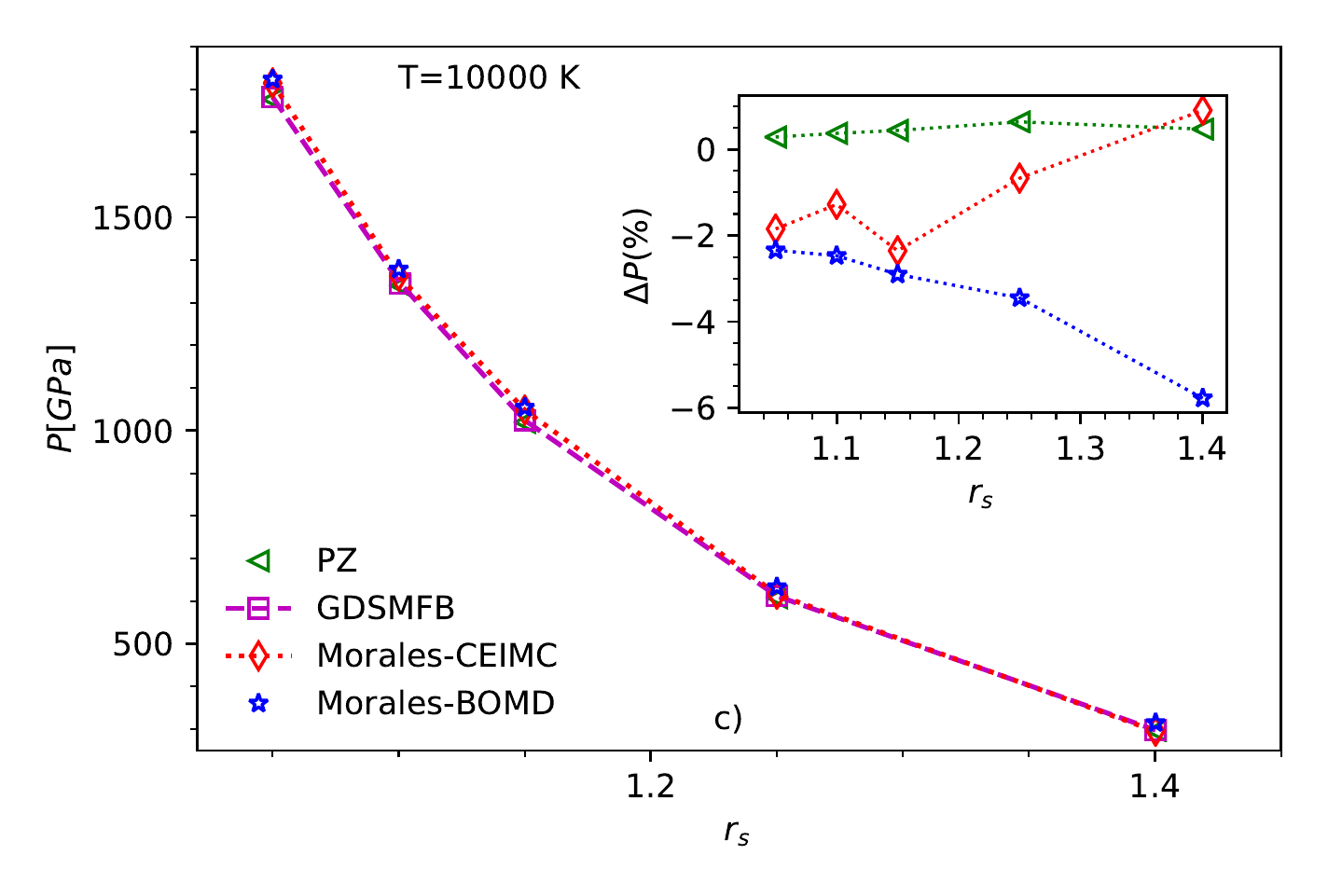}   
\caption{\raggedright EOS at a) $T=6000 K$, b) $T=8000 K$, and c) $T=10000 K$ comparing our results with the previous results obtained using PIMC and DFT. The inset plot shows the relative difference (Eq. \ref{pressure_eq1}) in pressure with respect to the finite temperature case.  Morales Ref.~\onlinecite{PhysRevE.81.021202}; Vorberger Ref.~\onlinecite{PhysRevB.75.024206}; Mazzola Ref.~\onlinecite{PhysRevLett.120.025701}. }  
\label{H2_6000}          
\end{figure}



Since LDA fails to capture the molecular dissociation correctly, we do not discuss the effect of finite temperature xc on the liquid-liquid phase transition (LLPT). 
Instead, this section focuses on the finite-temperature effects on the equation of state (EOS) in the high pressure fluid regime and a comparison is made to the available \textit{ab initio} results. 
\par 
We have performed simulations for a system size $N=256$ and the \textit{k}-point sampling is performed only at the $\Gamma$-point. While nuclear quantum effects (NQE) have been shown to influence the EOS at lower temperatures and the LLPT transition~\cite{Morales12799, morales2013nuclear}, they are not considered in this work as we focus on thermal xc effects on the electrons. Figure \ref{H2_1000} shows the EOS at 1000 K computed with various xc functionals and \textit{ab initio} methods. Here, the LLPT can be recognized by its characteristic signature in the PVT diagram~\cite{Morales12799} $(\partial P/\partial \rho )|_{T} = 0 $. CEIMC gives a very clear signature of the LLPT. The DFT-MD results, which can be calculated for a finer grid of points and have to utilize a dense \textit{k}-point grid near the transition region, show a slightly lower transition pressure~\cite{PhysRevB.82.195107}. 

First and foremost, we find no significant change in the EOS due to the incorporation of finite-temperature exchange and correlation effects as the reduced temperatures are low at these densities, $ 0.003 < \Theta < 0.004$. At $T=1000\ K$, the LDA results (PZ/GDSMFB) are in the range obtained by Alavi \textit{et al.}~\cite{Alavi1252}. There are obviously differences to the pressure isotherms obtained with different xc functionals, in particular in the molecular region. It is well known, that LDA cannot describe molecules properly. 
The CEIMC method for the EOS is in principle the most accurate under these conditions, and the next best approach would be DFT-MD if higher rungs of xc functionals, especially non-local density functionals, are used. Indeed, several papers recently reported a higher transition pressure comparable to and even above CEIMC when using functionals featuring van-der-Waals corrections~\cite{Knudson1455,Lu_2019}. 
 

In Figs. \ref{H2_6000}a-\ref{H2_6000}c, we compare the computed EOS with QMC and BOMD at temperatures 6000 K, 8000 K and 10000 K, respectively. The relative difference in pressure shown in the inset plot is given by 
\begin{equation}
\frac{(P_{GDSMFB}-P)}{P_{GDSMFB}} \times 100\%.  
\label{pressure_eq1} 
\end{equation}

Our LDA results and the BOMD results obtained by Morales \textit{et al.} using PBE  consistently exhibit deviations from each other with smallest deviations for the highest densities (lowest $r_{s}$) where the system is metallic~\cite{PhysRevE.81.021202}. 
While we expect CEIMC to be in principle the most accurate method, we observe a divergence between the latter and PBE with increasing $r_{s}$, where the CEIMC data more closely agree with our LDA results~\cite{PhysRevE.81.021202}. One possible explanation for this trend would be a systematic bias in the CEIMC data due to the employed LDA-based trial wave function in this approach, although this cannot be resolved on the basis of our data.
In Fig. \ref{H2_6000}a, we also show the close agreement for lower densities ($r_{s}>1.4$) between our results and QMC as well as PBE-BOMD data obtained by Mazzola \textit{et al.}~\cite{PhysRevLett.120.025701} and Vorberger \textit{et al.}~\cite{vorberger_hydrogen-helium_2007}. The GDSMFB results in a higher pressure compared to PZ by $0.2-0.5 \%$ at these conditions, which is reasonable as $\Theta \approx 0.01-0.03$. A similar trend can also be observed in the Fig. \ref{H2_rs_0p7_P_vs_T} discussed in section \ref{wdm} where the pressure difference is positive for similar densities at low temperatures.          
\par 
For completeness, we show the phase diagram with LLPT boundaries in Fig. \ref{H2_phase_diagram} in Appendix~\ref{sec:appendix_phase}.

\subsection{Warm dense matter}  
\label{wdm}

Let us next explore the warm dense matter regime, where thermal xc effects are expected to be important.  
 
\par
\begin{figure}[H]           
\centering    
\includegraphics[width=1.0\columnwidth]{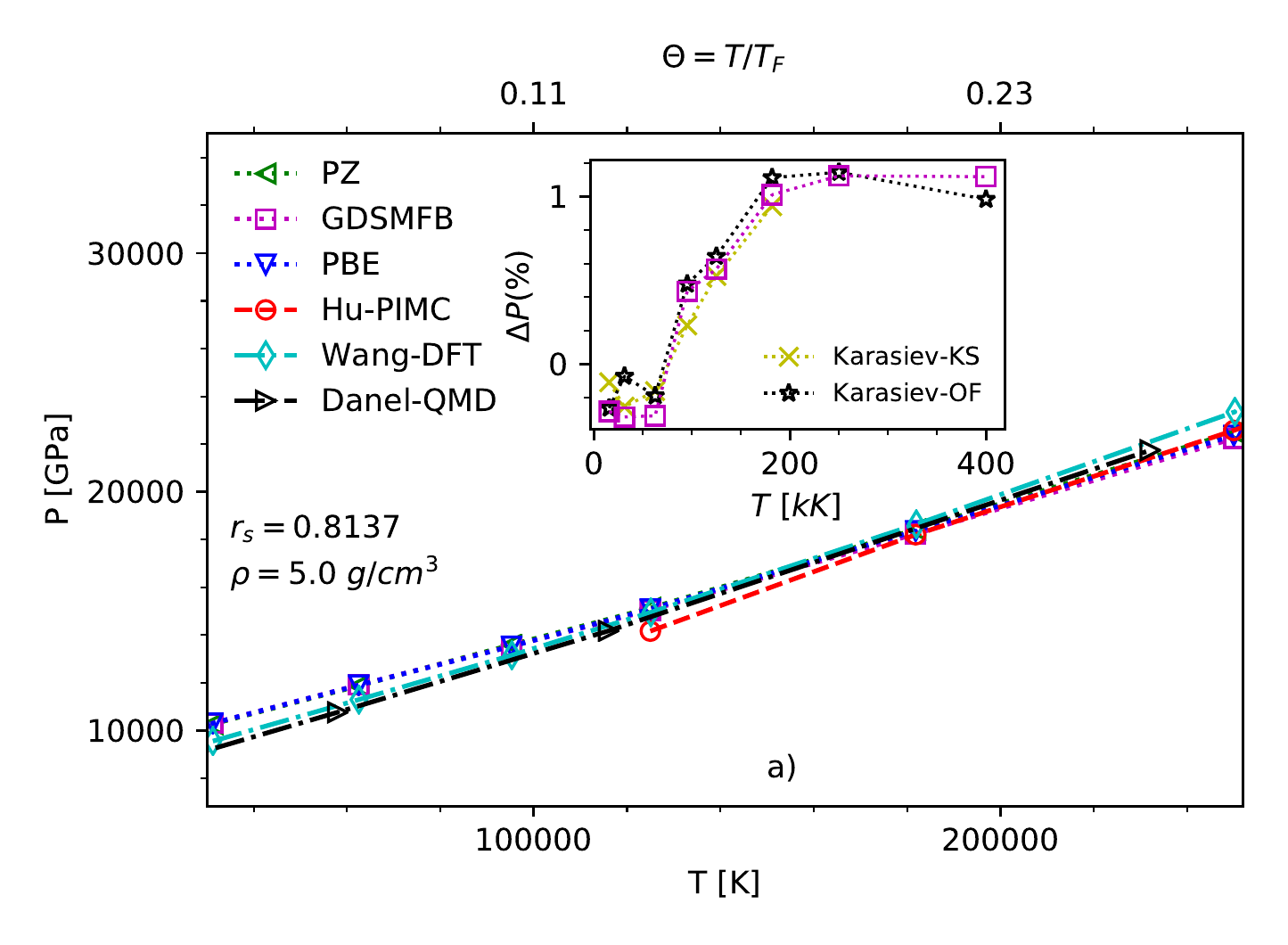}
\includegraphics[width=1.0\columnwidth]{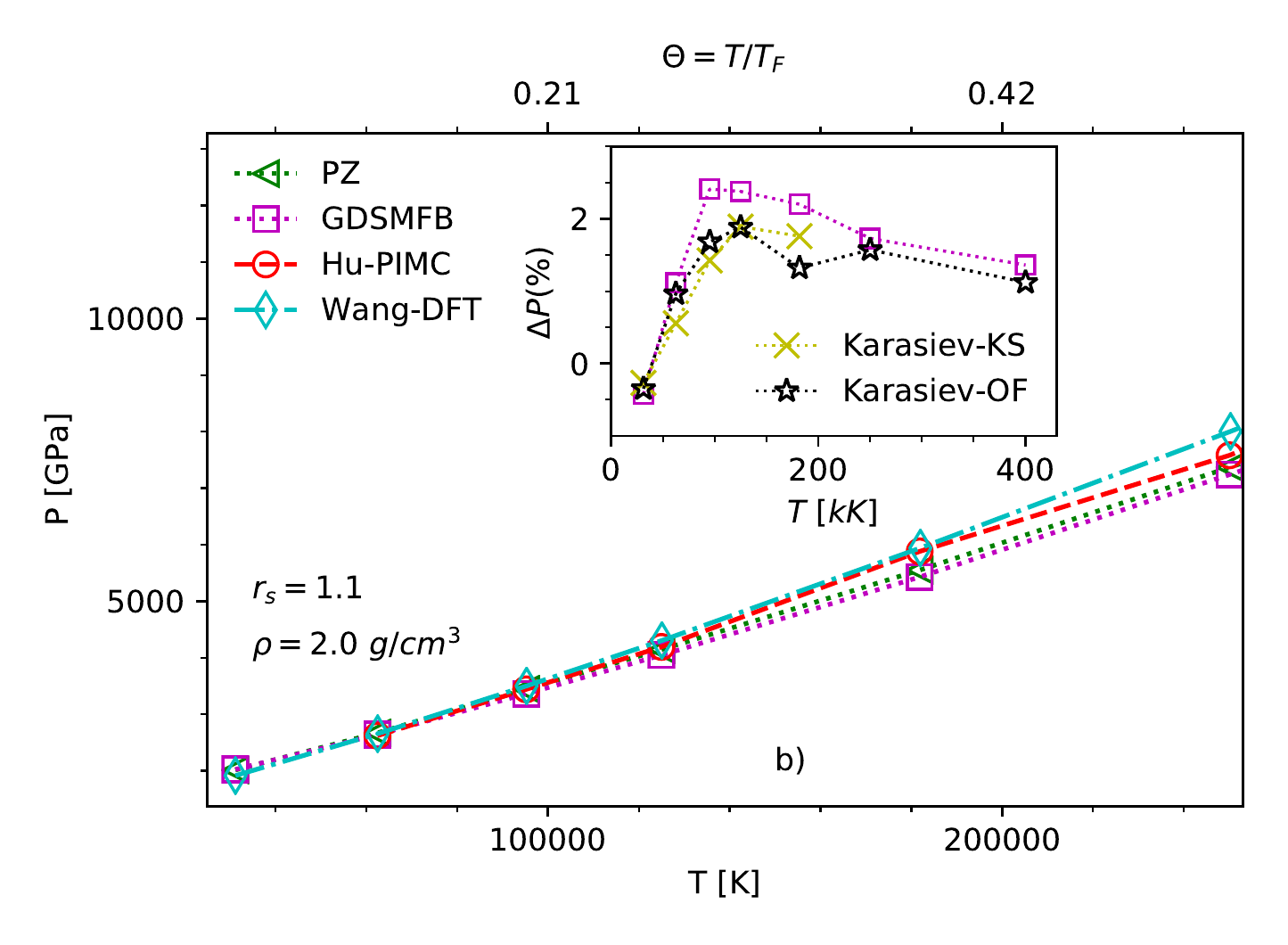}
\includegraphics[width=1.0\columnwidth]{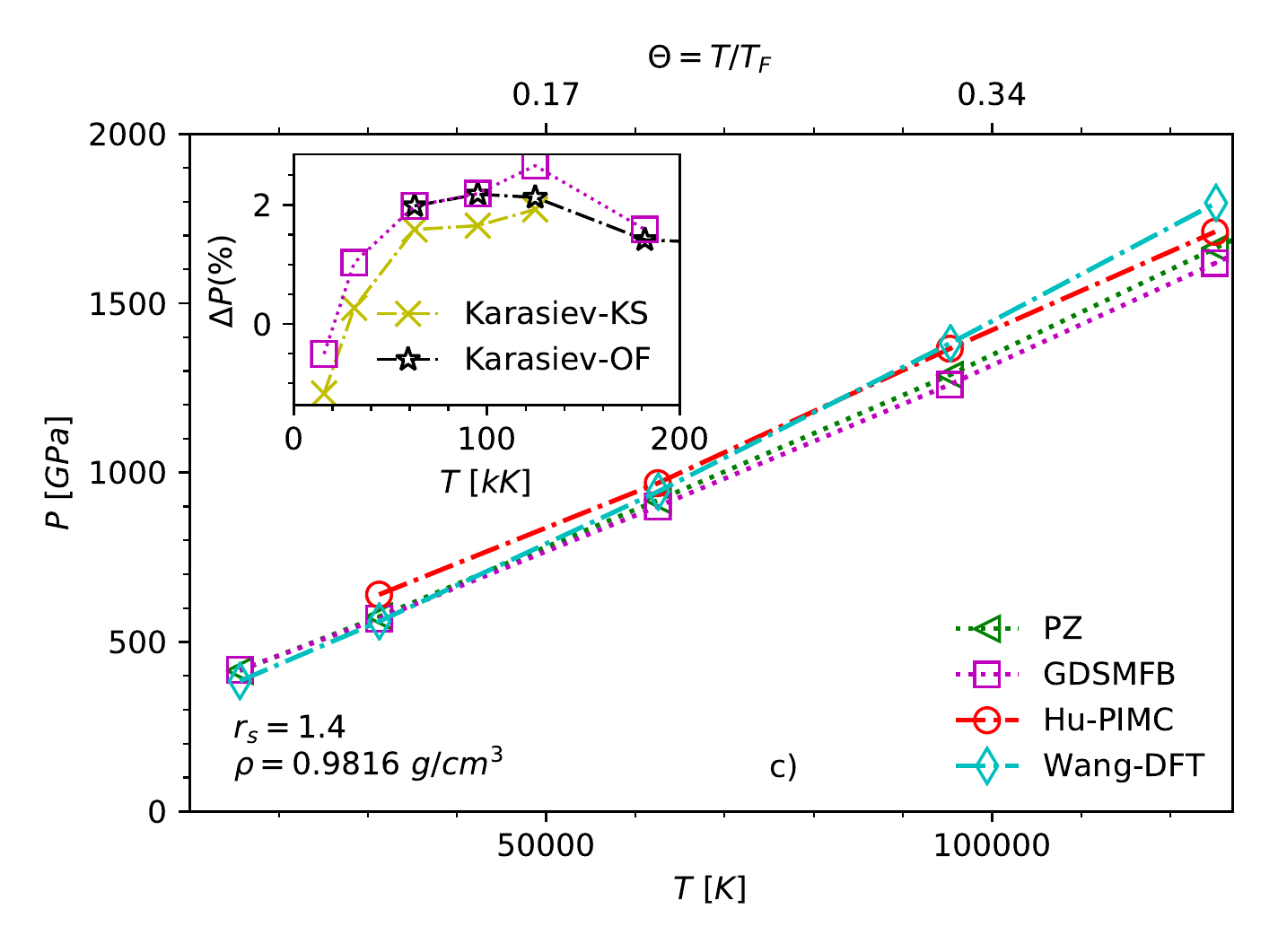}
\caption{\raggedright EOS at a) $r_{s} = 0.8137$, b) $r_{s} = 1.1$, c) $r_{s} = 1.4$ comparing our results with the previous results obtained using PIMC and DFT. The inset plot shows the relative difference (Eq. \ref{pressure_eq2}) in total pressure for the finite temperature case with respect to the LDA case. Hu Ref.~\onlinecite{PhysRevB.84.224109}; Wang (LDA) Ref.~\onlinecite{doi:10.1063/1.4821839}; Danel (LDA) Ref.~\onlinecite{PhysRevE.93.043210}; Karasiev Ref.~\onlinecite{PhysRevE.93.063207}. }       
\label{H2_rs_0p7_P_vs_T}         
\end{figure}

In Figs. \ref{H2_rs_0p7_P_vs_T}a-\ref{H2_rs_0p7_P_vs_T}c, we compare the equation of state for $r_s=0.8137\ldots 1.4$ corresponding to densities in the range of $5.0\ldots 0.98\,$g/cm$^{3}$. Overall, the agreement between LDA-BOMD of different sources and PIMC is reasonable. Our results are consistently closest to PIMC of all DFT-MD data at the highest temperatures. However,
our results for the pressure deviate from the data by Wang \emph{et. al}~\cite{doi:10.1063/1.4821839} at lower temperatures for $r_{s}=$0.8137 as can be seen in Fig. \ref{H2_rs_0p7_P_vs_T}a.
This is due to the increased computational cost of the mixed Gaussian plane wave method at high densities and  temperatures that forced us to choose $N=32$ for all of the simulations shown in this section. Moreover, the \textit{k}-point sampling is performed only at the $\Gamma$-point. The resulting finite-size effects have been demonstrated for lower temperatures at $r_s=3$, where we have obtained higher pressures for smaller system sizes. Wang \textit{et al.}~\cite{doi:10.1063/1.4821839} use a system size ranging from 8-512 atoms sampled at the $\Gamma-$point with the Kohn-Sham DFT simulations restricted to $T<T_{F}$ for $\rho>0.5 \ g/cm^{3}$. Yet, this does not necessarily constitute a problem, since we are only interested in the pressure differences due to the use of the GDSMFB functional instead of PZ, and the finite-size effects are expected to mostly cancel.  
For example, a similar cancellation of finite-size effects has been reported for \textit{ab initio} PIMC calculations of the static local field correction, see, e.g., Refs.~\onlinecite{dornheim2019static,groth_jcp17}.


Thus, the relative difference in the total pressure calculated using
\begin{equation}
\frac{(P_{PZ} - P_{GDSMFB})}{P_{PZ}} \times 100\%.  
\label{pressure_eq2} 
\end{equation}
shown in the inset plots in Fig. \ref{H2_rs_0p7_P_vs_T} is in good agreement with the Kohn-Sham DFT and orbital-free DFT results obtained by Karasiev \textit{et al.}~\cite{PhysRevE.93.063207}. Danel \textit{et al.}\cite{PhysRevE.93.043210} estimated the temperature dependence in the xc functional using a fit by Ichimaru. Their results are not compatible with ours.


\par
\begin{figure}[th]     
\centering  
\includegraphics[width=1.0\columnwidth]{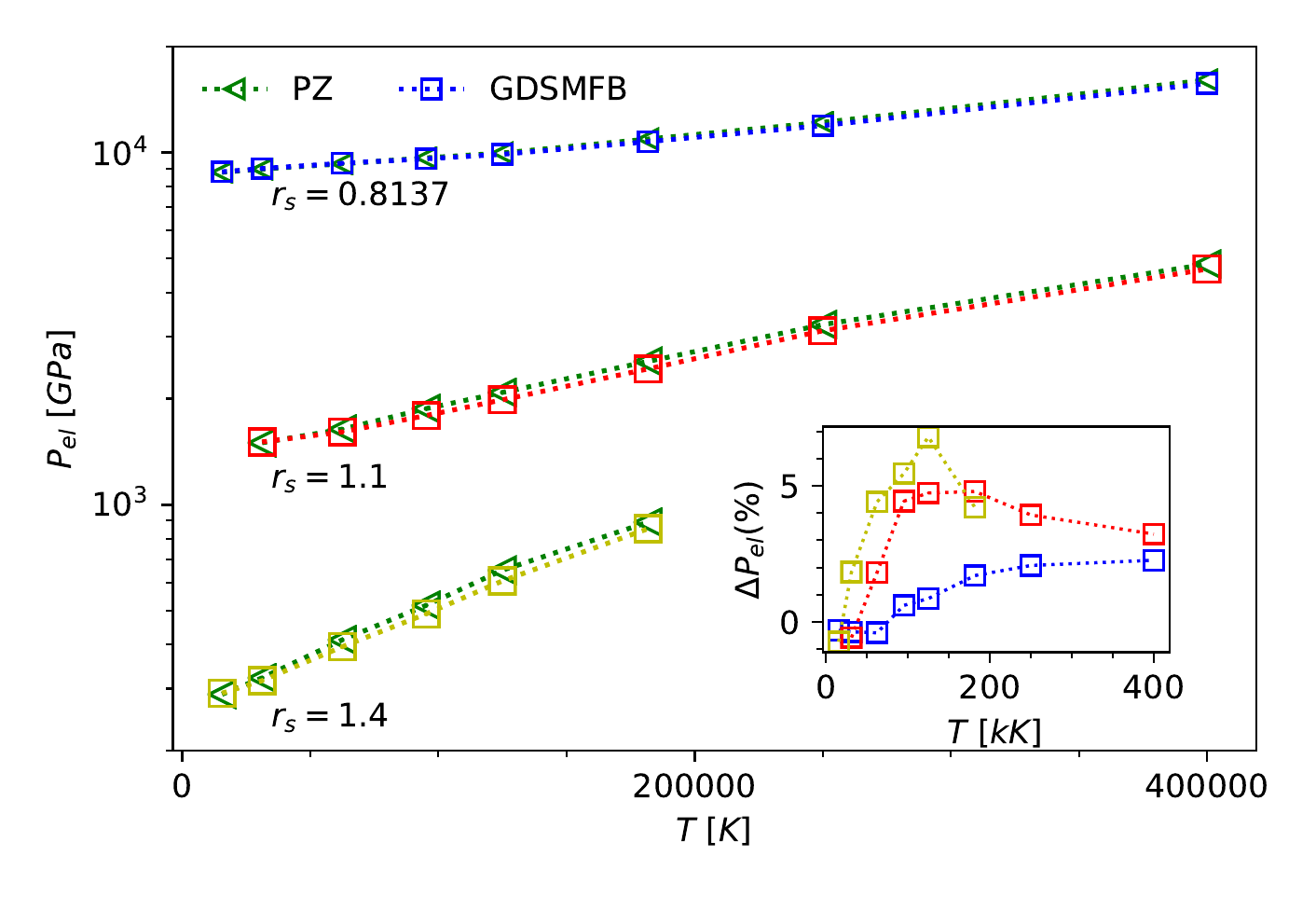}  
\caption{\raggedright The electronic pressure as a function of temperature at various densities. The inset plot shows the relative difference (Eq. \ref{pressure_eq2}) in electronic pressure for the finite temperature case with respect to the LDA case. } 
\label{H2_electronic_P}           
\end{figure}

  
The variation of the electronic pressure with respect to temperature and density is shown in Fig. \ref{H2_electronic_P}. In accordance with Ref.~\onlinecite{PhysRevE.93.063207}, we obtain the electronic pressure by subtracting the ideal ion pressure from the total pressure. The ionic excess pressure as can be obtained for instance by integrating over the pair correlation function, has not been subtracted~\cite{kremp2006quantum}. With a decrease in density, the relative difference in electronic pressure at a fixed temperature increases as the temperature effects on the electronic correlations are more prominent as the Fermi temperature decreases with density. At 125000 K, as the density decreases from $r_{s}=0.8137$ to 1.4, $\Theta$ changes from 0.14 to 0.42 and a large deviation in the electronic pressure is noted. At $r_{s}=0.8137$, notable deviations in the electronic pressure begin to appear at temperatures above 400000 K which, however, is beyond the scope of this work based on Kohn-Sham DFT.

Summarizing, our new simulation results further corroborate the observations by Karasiev \textit{et al.}~\cite{PhysRevE.93.063207}, and stress the importance of finite-$T$ xc effects for DFT simulations in the WDM regime.

\subsection{Moderately coupled plasma and electron liquid regime}
 
\begin{figure}[th]    
\centering 
\includegraphics[width=1.0\columnwidth]{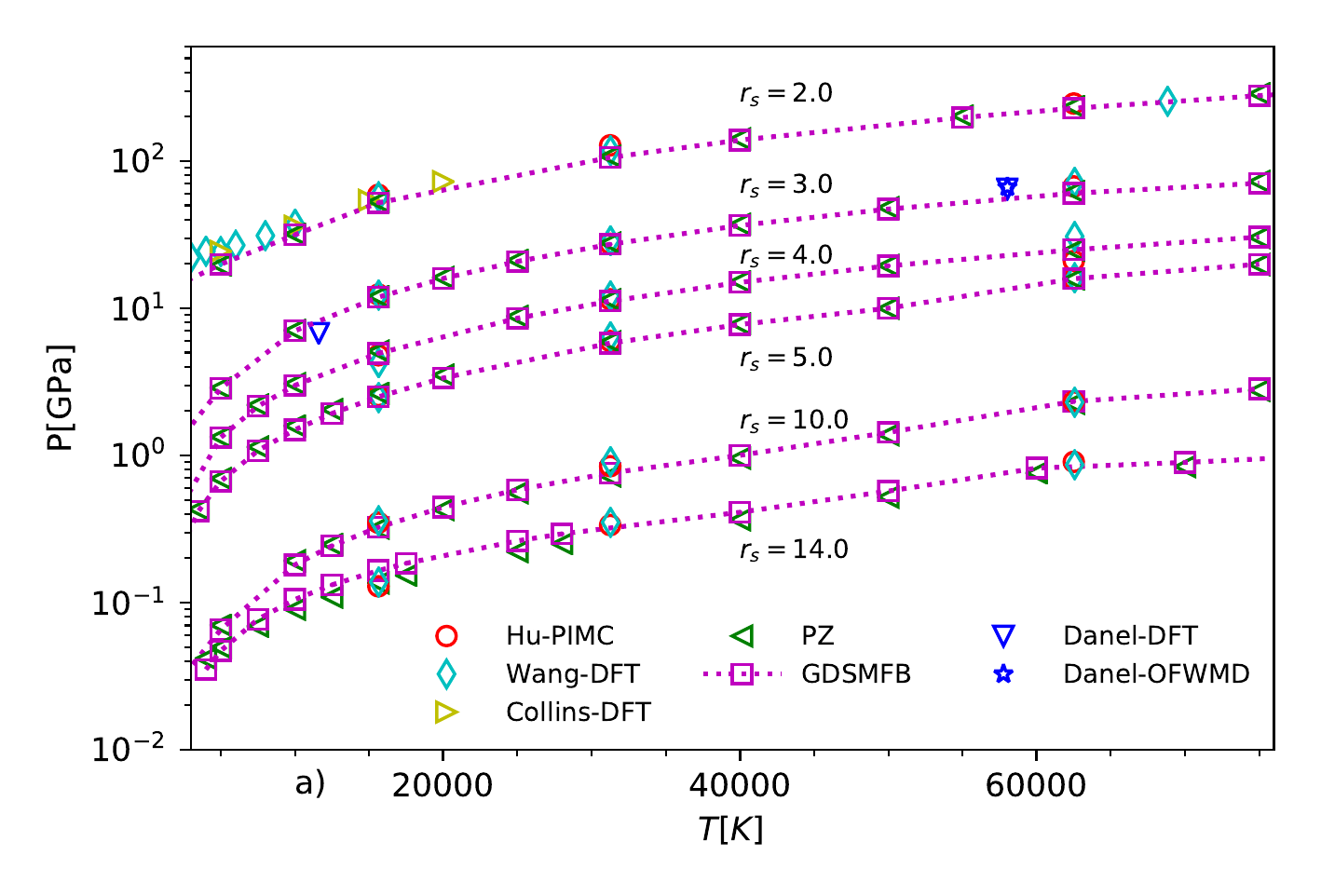} 
\includegraphics[width=1.0\columnwidth]{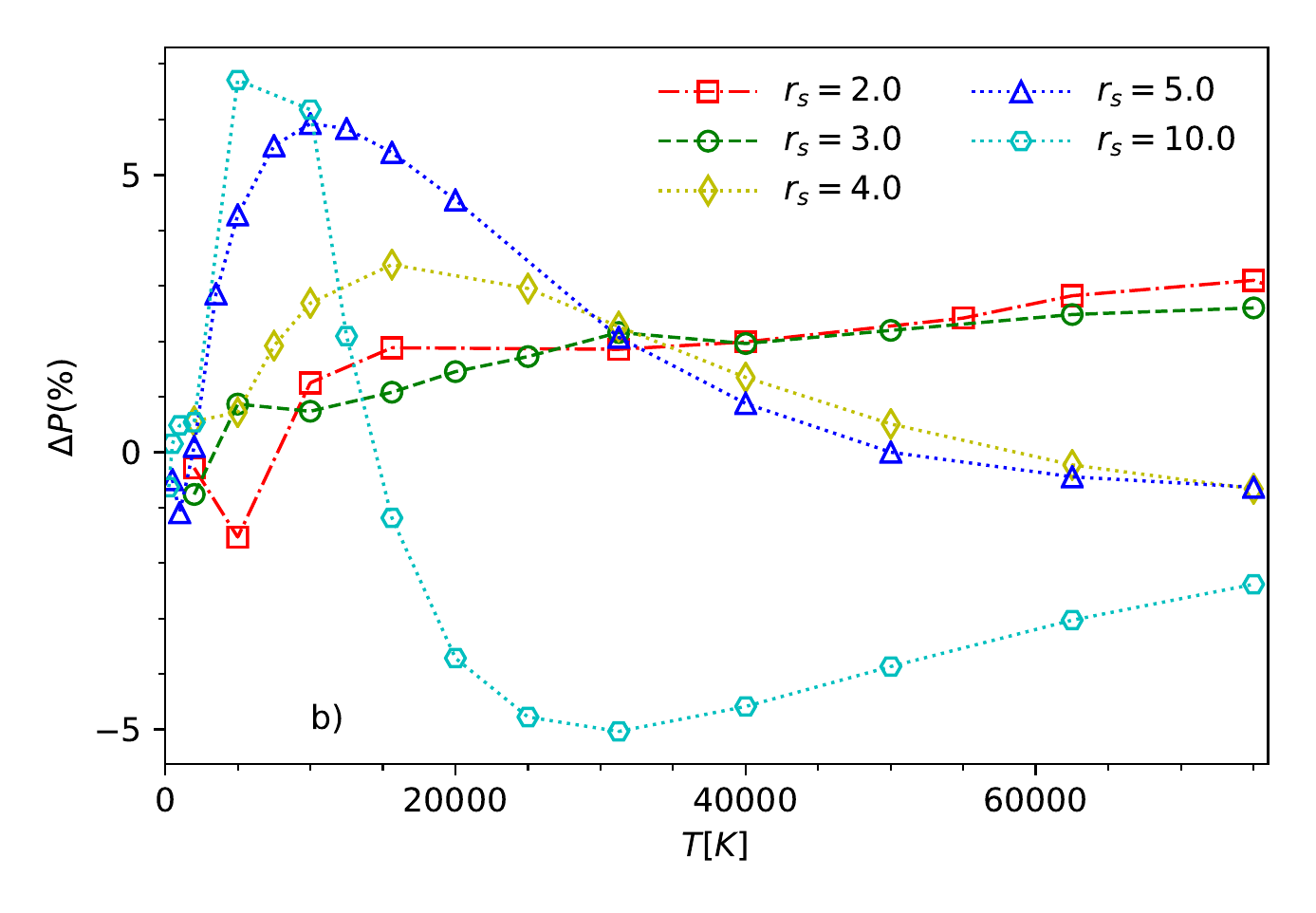} 
\caption{\raggedright a) EOS at $r_{s}=2.0-14.0$ comparing our results with the previous results obtained using PIMC and DFT. b) Relative difference (Eq. \ref{pressure_eq2}) in total pressure for the finite temperature case with respect to LDA at $r_{s}=2.0-10.0$. Hu Ref.~\onlinecite{PhysRevB.84.224109}; Wang Ref.~\onlinecite{doi:10.1063/1.4821839}; Collins Ref.~\onlinecite{PhysRevB.63.184110}; Danel Ref.~\onlinecite{PhysRevE.93.043210} } 
\label{H2_EOS_rs_2_10}              
\end{figure}

\begin{figure}[th]   
\centering    
\includegraphics[width=1.0\columnwidth]{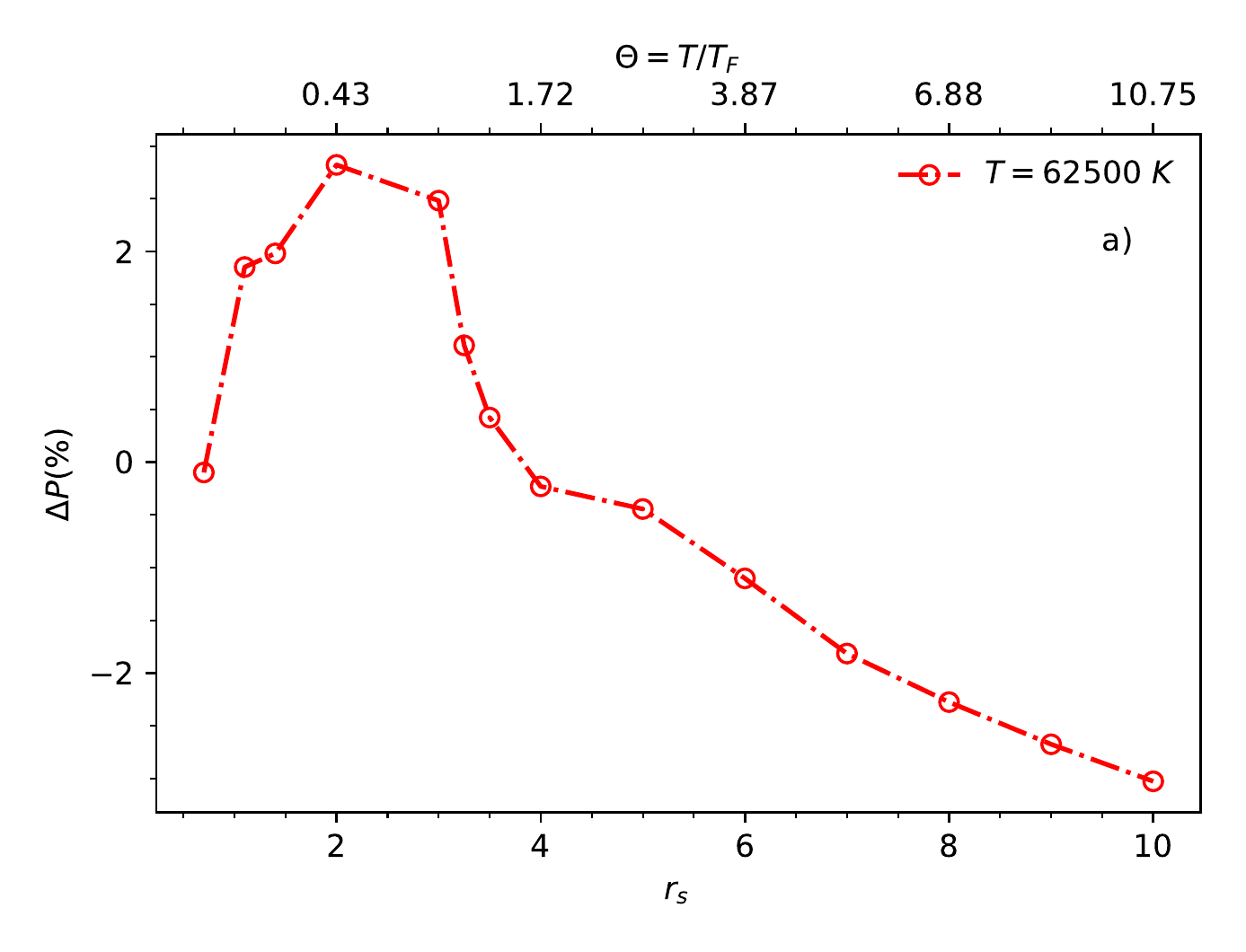} 
\includegraphics[width=1.0\columnwidth]{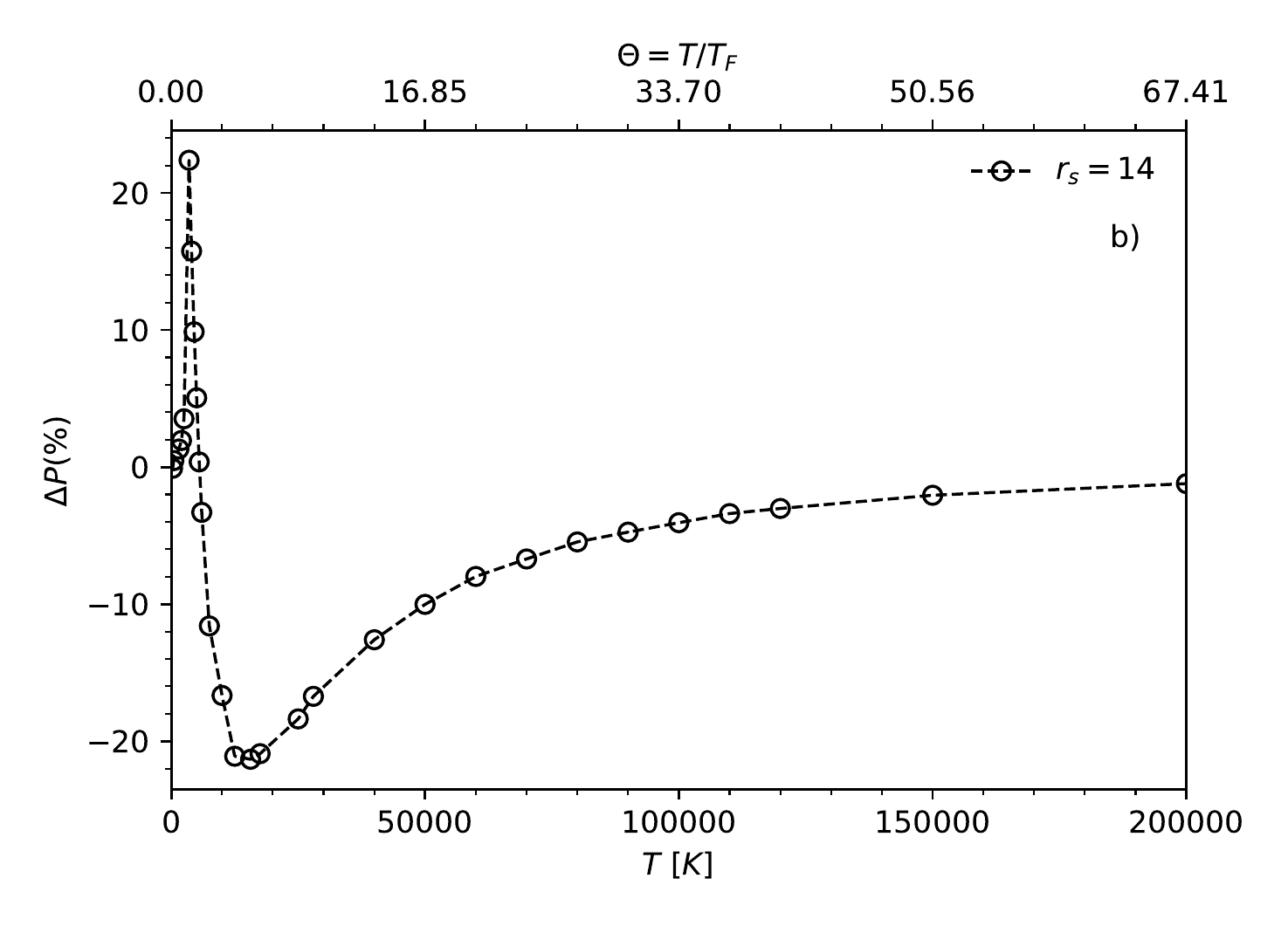}  
\caption{\raggedright a) Relative difference (Eq. \ref{pressure_eq2}) in total pressure for the finite temperature case with respect to LDA with the change in density ($\leftarrow r_{s}$) and electron degeneracy at $T=62500 \ K$; b) Relative difference in total pressure for the finite temperature case with respect to LDA at $r_{s}=14$.} \label{H2_p_vs_rs_62500K}                
\end{figure}        
 

With decreasing density (i.e., increasing $r_s$), electronic correlations become more important and the system will eventually form an electron liquid~\cite{giuliani_vignale_2005,PhysRevB.24.7385,PhysRevB.24.3220,dornheim_prl_2018,groth_prb_2019} for $r_s\gtrsim10$. 3D electron liquids can be found in metals 
if the Fermi surface of the conducting electrons is spherical 
to facilitate the movement of electrons as free particles. This is also possible with semiconductors by controlling the amount of dopants~\cite{giuliani_vignale_2005,PhysRevB.94.245106}. While these exotic conditions are rather difficult to realize experimentally at present, they offer the valuable opportunity to study the nontrivial interplay of temperature and Coulomb coupling  ($\Gamma=1/(a k_{B} T_{e} )$) with quantum diffraction and exchange effects. For example, Takada~\cite{PhysRevB.94.245106} predicted the emergence of a collective excitonic mode for large $r_s$ based on ground-state many-body theory, which was recently substantiated by more accurate \textit{ab initio} path-integral Monte-Carlo calculations at finite temperature~\cite{dornheim_prl_2018,groth_prb_2019}. Moreover, the possibility of an experimental detection of the associated negative dispersion relation of the dynamic structure factor constitutes an exciting opportunity for future research~\cite{dornheim_prl_2018}.


\par 
Figure \ref{H2_EOS_rs_2_10}a shows the EOS for $r_s=2.0\ldots 14$ corresponding to densities in the range $9.8 \times 10^{-4}\ldots 0.34\,$g/cm$^{3}$.  The system size is $N=256$ except at $r_s \geq 10 $ where the system size is reduced to $N=32$ due to the large simulation box required at extremely low densities. The \textit{k}-point sampling is performed only at the $\Gamma-$point. The EOS fits well with the PIMC and DFT-MD data of Hu \textit{et al.} and Wang \textit{et al.}, respectively across a gamut of temperatures for the densities considered~\cite{PhysRevB.84.224109,doi:10.1063/1.4821839}. The relative difference in total pressure between PZ and GDSMFB is shown in Fig. \ref{H2_EOS_rs_2_10}b.

First and foremost, we note that $\Delta P$ exhibits a sign change for intermediate temperatures, which is shifted to larger temperatures for increasing density. This is again a consequence of the $r_s$-dependence of the reduced temperature $\Theta$, which decreases with $r_s$. Such a sign change has been reported for the pressure in previous DFT calculations in Refs.~\onlinecite{PhysRevE.86.056704,PhysRevB.90.155109,PhysRevE.93.063207}. Moreover, a similar behaviour was found in QMC calculations for the xc-part of the kinetic energy of the UEG, see, e.g., Ref.~\onlinecite{PhysRevLett.89.280401}.

In the range $r_s=5\ldots 10$, the relative difference in pressure is more pronounced with positive differences at low temperatures and negative differences at higher temperatures being of a similar magnitude. The maximum changes for $r_s=10$ are observed in a broad range of reduced temperatures of $\Theta=0.6-6.0$. For comparison, we mention that the positive maximum deviation for $r_s=5$ is found for $\Theta=0.3-0.7$, whereas the negative maximum deviation extends to temperatures beyond the depicted range. At $r_s=2.0\ldots 3.0$, the onset of the significant changes begin near the maximum of the temperature considered in Fig. \ref{H2_EOS_rs_2_10}b. This can be observed in Fig. \ref{H2_p_vs_rs_62500K}a, where the temperature is held constant and the relative difference in total pressure is evaluated with the change in density and the electron degeneracy. The positive pressure difference is maximal for the density range $r_s=2.0\ldots 3.0$ in the vicinity of $\Theta \sim 1$ and $\Gamma \sim 2$.

For very low densities, e.g., $r_s=14$, the plane wave energy cutoff needs to be increased to 800 Ry in order to achieve convergence. Then, agreement with Hu \textit{et al.} and Wang \textit{et al.}, respectively, across a range of temperatures is given~\cite{PhysRevB.84.224109,doi:10.1063/1.4821839}. See appendix \ref{app_high_rs} for an investigation of the convergence with respect to the energy cutoff. The system size is set to $N=32$ due to the large simulation box and sampled only at the $\Gamma-$point. As before, finite size effects should be unimportant as we are only interested in the relative differences in total pressure.    

At $r_s=14.0$, we compute the relative difference in total pressure across a wide range of temperature where the maximum relative differences can be seen at the reduced temperatures $\Theta=1.18$ and $\Theta=5.90$, respectively, as shown in Fig. \ref{H2_p_vs_rs_62500K}b. Remarkably, these deviations exceed $20\%$ and, thus, are even more pronounced than for the previously considered WDM regime.

\subsection{Density of states} 


\begin{figure*}[th]     
\centering    
\includegraphics[width=1.0\columnwidth]{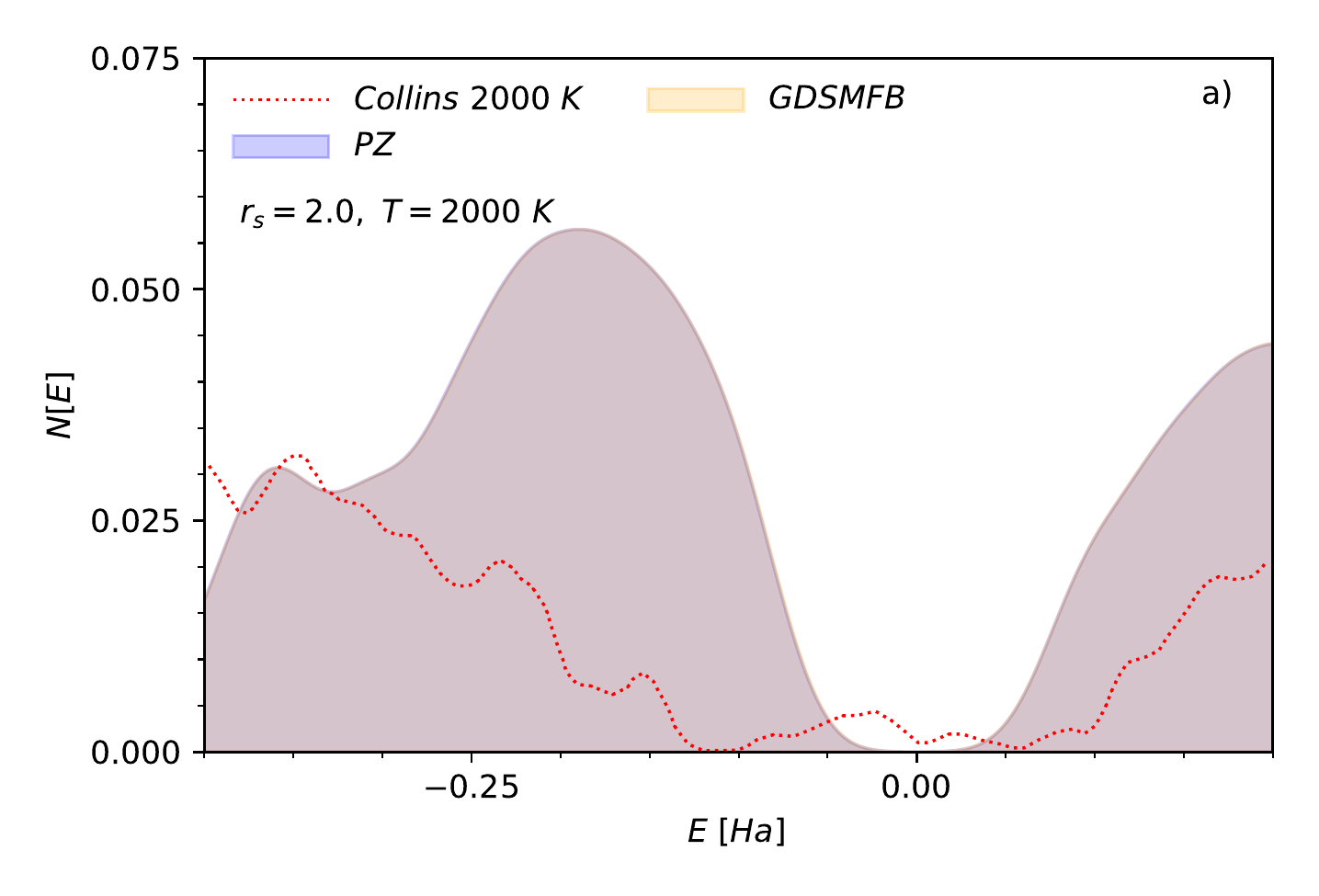} 
\includegraphics[width=1.0\columnwidth]{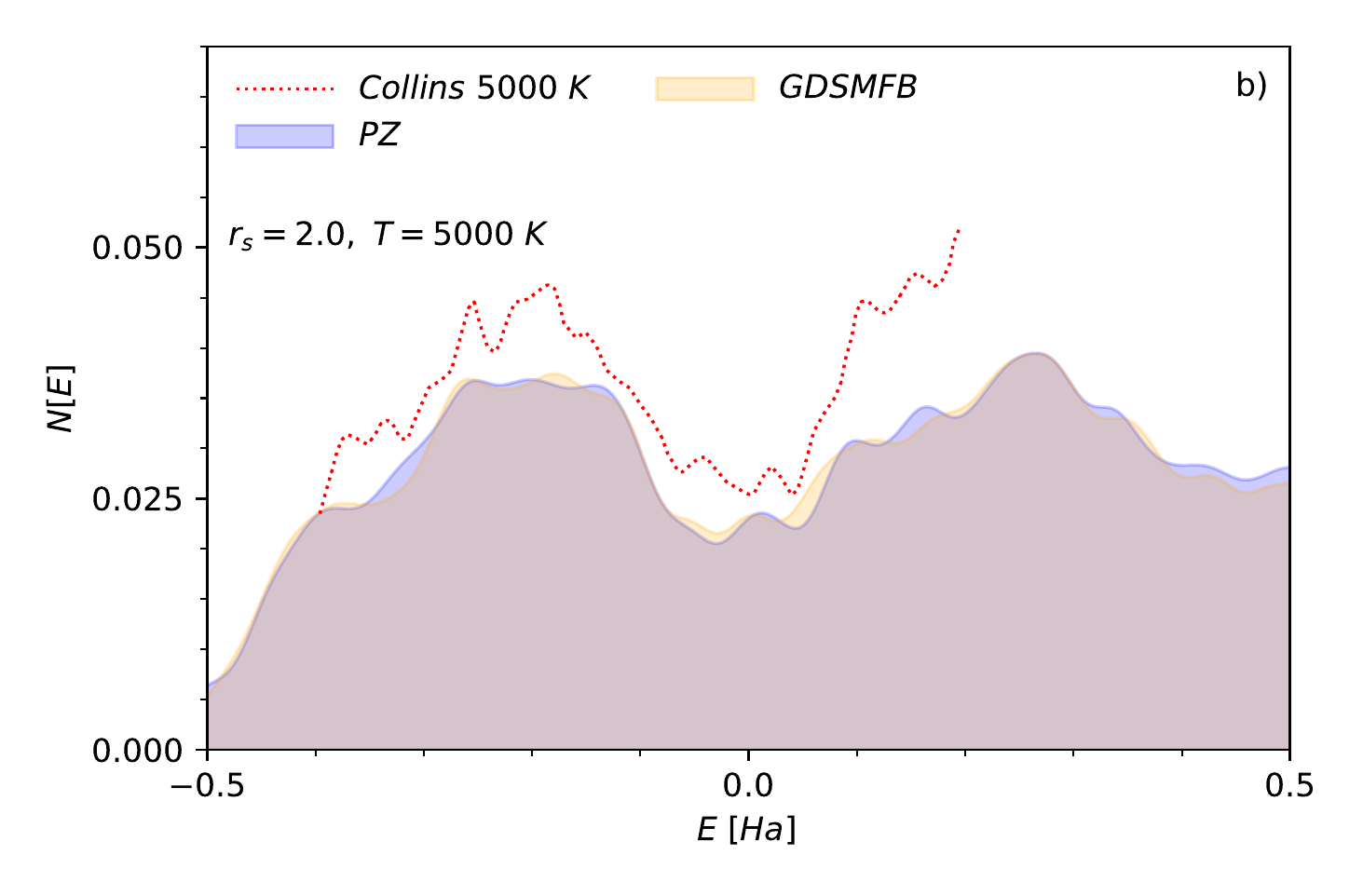}
\includegraphics[width=1.0\columnwidth]{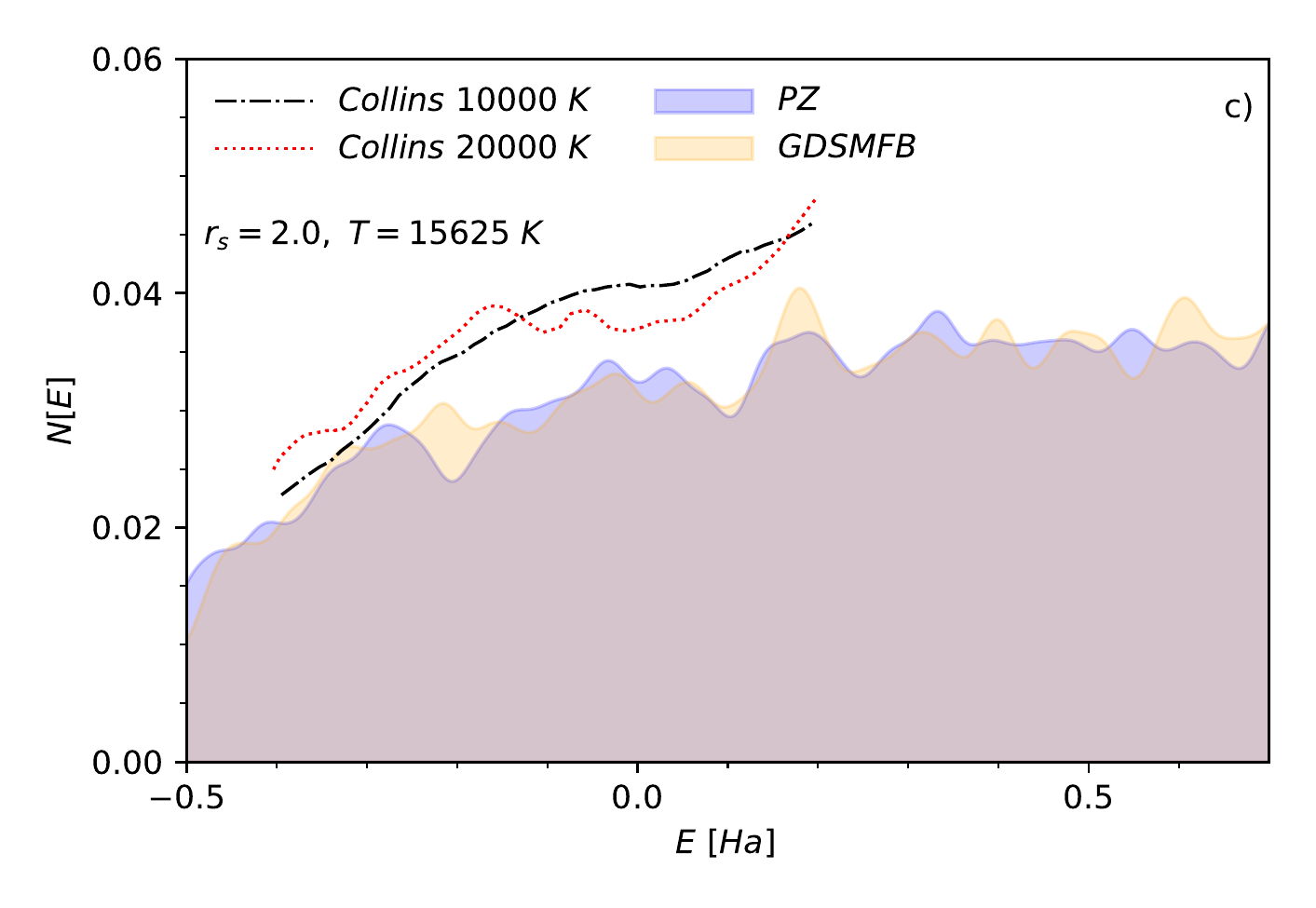}  
\includegraphics[width=1.0\columnwidth]{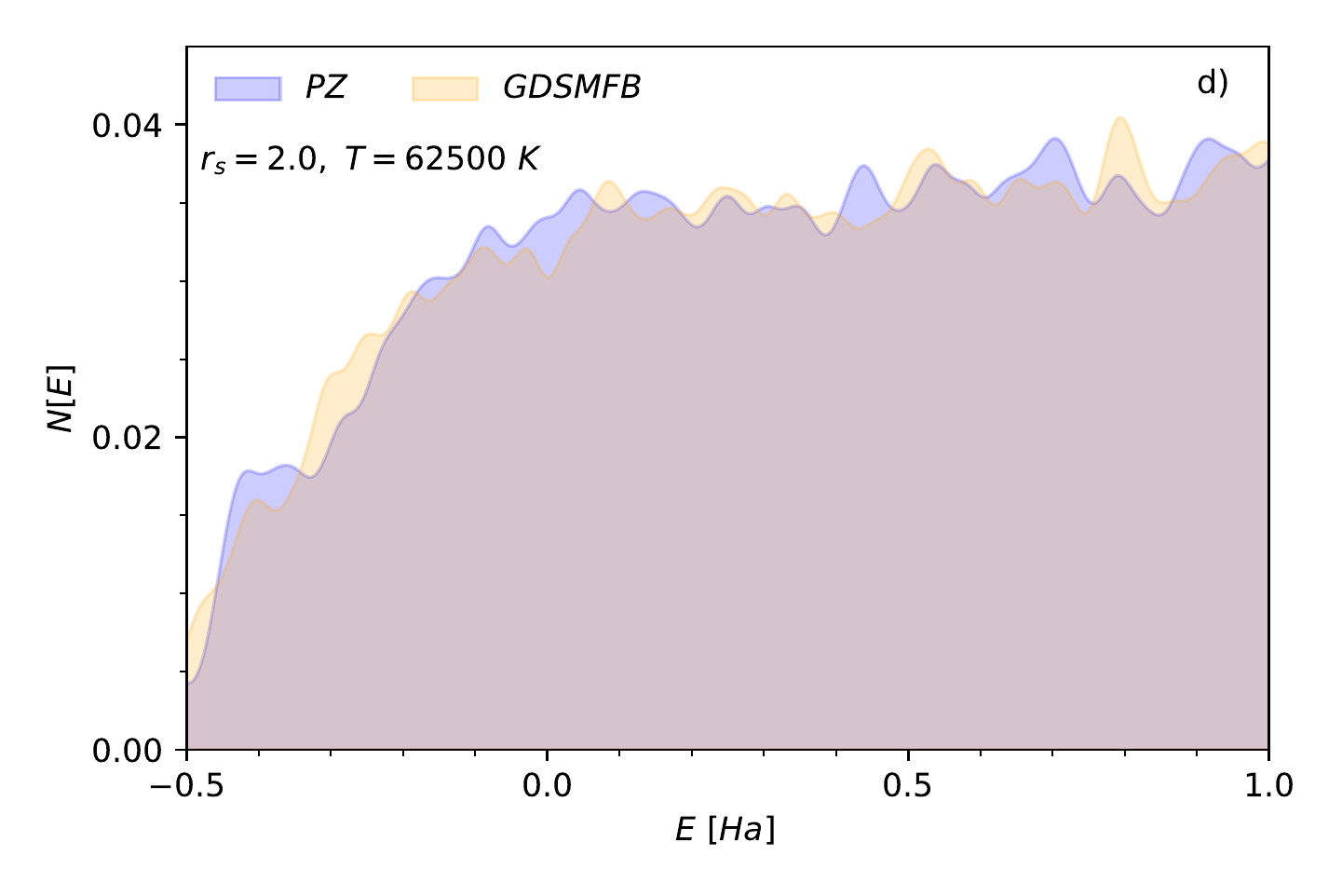}  
\caption{\raggedright The density of states at $r_{s} = 2.0$ and a) $T=2000$ K b) $T=5000$ K c) $T=15625$ K and d) $T=62500$ K using LDA and the finite-temperature case. The Fermi energy is set to zero. The blue and the yellow areas represent DOS calculated using PZ and GDSMFB respectively. The overlap between them is indicated by the violet area.  
Collins Ref.~\onlinecite{PhysRevB.63.184110}. }  \label{H2_DOS_rs_2_15625K}  
\end{figure*}




  
The density of states (DOS) is computed for $r_{s}=2.0$ simulating N$=256$ atoms sampled at the $\Gamma$-point. We choose a set of 5 independent equilibrated configurations from different simulation runs which are averaged to obtain the corresponding DOS. In Figs. \ref{H2_DOS_rs_2_15625K}a-\ref{H2_DOS_rs_2_15625K}d, the DOS is shown for a range of temperatures and compared to the ground-state GGA calculations by Collins \textit{et al.}~\cite{PhysRevB.63.184110}. At $T=2000$ K, the system is still insulating with a band gap shown in Fig. \ref{H2_DOS_rs_2_15625K}a while the results from Collins \textit{et al.} show a slight increase in the DOS near the Fermi level. At a slightly higher temperature of $T=5000$ K, the system is metallic and our results match the trend obtained by  Collins \textit{et al.} Between $T=2000$ and $T=5000$ K, the DOS is hardly influenced by finite temperature xc-effects as the reduced temperature is still low. At $T=15625$ K ($\Theta=0.107$), the results follow the trend seen by  Collins \textit{et al.}, with the $\sqrt{E}$ feature being clearly visible at 62500 K ($\Theta=0.43$). Noticeable differences between the DOS computed with the PZ T$=$0 functional and the GDSMFB finite-$T$ functional start to appear at these two temperatures, which are still below the regime where the maximum change in finite temperature xc effects can be observed.

\subsection{Electronic density}
 
\begin{figure*}[th]             
\centering  
\includegraphics[width=0.56\columnwidth]{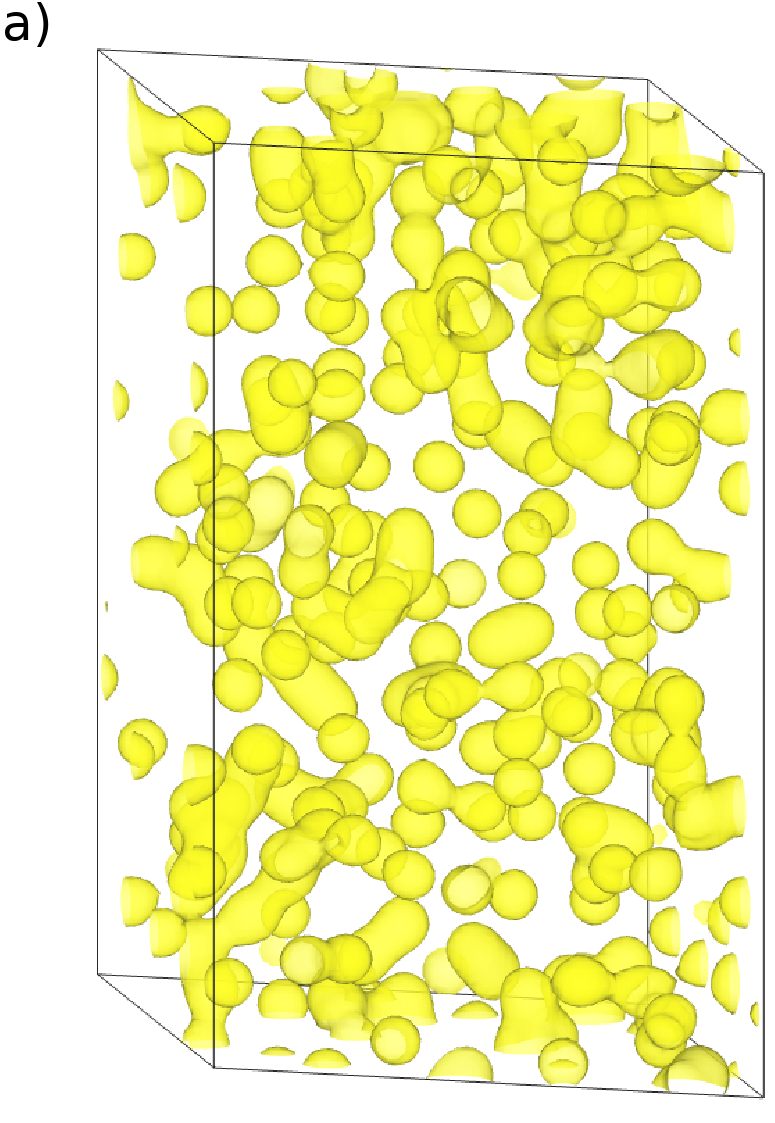}
\includegraphics[width=0.56\columnwidth]{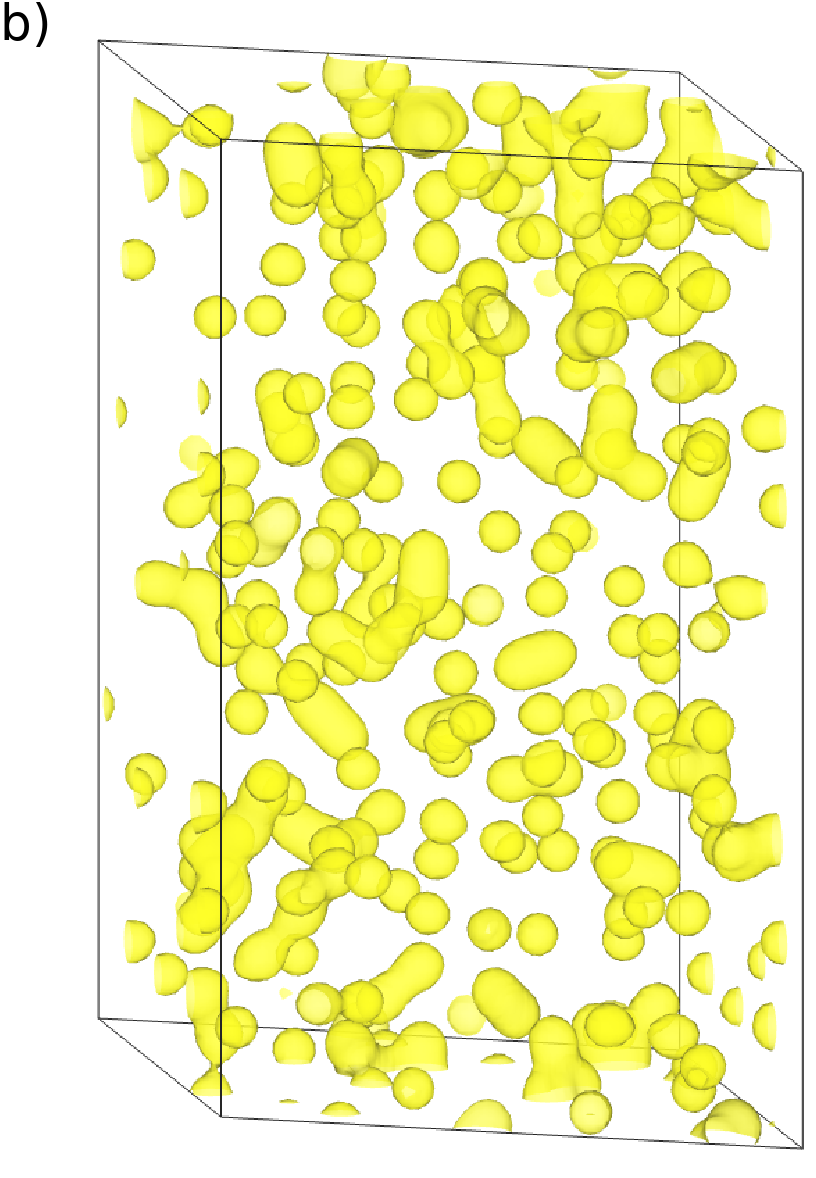}\hspace*{0.431cm}
\includegraphics[width=0.375\columnwidth]{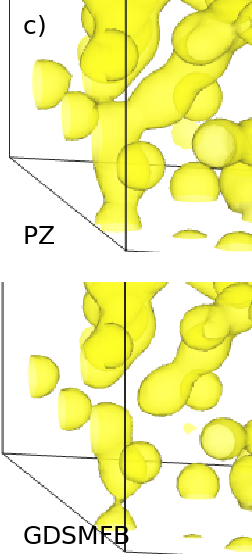}
\caption{\raggedright Snapshot of an electronic density isosurface for $r_s=2.0$, $T=$ 62500 K ($\Theta=0.43$) using a) PZ and b) GDSMFB for the same ionic configuration with $N=256$. Panel c) shows magnified insets for PZ (top) and GDSMFB (bottom) for the bottom left corner of the respective simulation cells.  }
\label{H2_rs_2_el_density_snapshot3}
\end{figure*}   


Figure \ref{H2_rs_2_el_density_snapshot3}a-\ref{H2_rs_2_el_density_snapshot3}b shows snapshots of an electronic density iso-surface computed using PZ and GDSMFB for $r_{s}=2.0$, $T=62500$ K (i.e., $\Theta=0.43$, which is located in the WDM regime) and for the same ionic configuration. These  stem from an $N=256$ hexagonal super cell sampled at the $\Gamma-$point. The snapshot has been obtained by performing DFT-MD simulations with the PZ functional until the system equilibrated and a random ionic configuration is chosen, which is subsequently used to compute the density with the different xc functionals. The visualization of the results are generated using VESTA~\cite{Momma:db5098}.          
\par   

Overall, the two snapshots in panels a) and b) exhibit a similar structure, with the electronic iso-surfaces being mostly located around the ions. Yet, there appear distinct systematic differences, which can be seen particularly well in Fig.~\ref{H2_rs_2_el_density_snapshot3}c) showing a magnified segment around the bottom left corner of the simulation cell: using the PZ-functional (top), there appears a pronounced overlap between the electronic orbitals around individual atoms; the GDSMFB-functional (bottom), on the other hand, leads to substantially reduced overlap, as we shall explain heuristically in the following. With increasing temperature, the thermal wavelength $\lambda\sim 1/\sqrt{T}$ decreases and, consequently, the electronic orbitals are less extended. Ultimately, this leads to the convergence to classical point-like particles in the high temperature limit.
The PZ-functional, which has been constructed solely based on ground-state data for the UEG, cannot consistently capture this behaviour, and the extension of the electronic iso-surfaces is drastically overestimated.

We thus conclude that including finite-$T$ xc effects in a thermal DFT simulation of a WDM system is crucial to capture the relevant physics, even though the impact on averaged quantities like the total pressure ($3\%$, cf.~Fig.~\ref{H2_p_vs_rs_62500K}) might be comparably small. The local electronic density and it's fluctuation are for instance important for the calculation of response functions, dielectric functions, and thus the prediction of structure factors as they are measured, e.g., via x-ray scattering~\cite{falk_2018,graziani2014frontiers}.

\section{Conclusions}

In summary, we have studied in detail the impact of finite-temperature xc effects on the results of DFT simulations of hydrogen over a vast range of different conditions. More specifically, we have carried out extensive DFT calculations using the ground-state functional by Perdew and Zunger (PZ) and the recent finite-$T$ analogue by Groth \textit{et al.}~\cite{PhysRevLett.119.135001} (GDSMFB). This has allowed us to unambiguously quantify the impact of finite-$T$ xc for different quantities and in different physical regimes.

Firstly, we have found that electronic temperature effects do not play a significant role for the description of the LLPT, as the reduced temperature is small, $\theta\lesssim0.01$. Thus, closing the gap between simulation and experiments will most likely require to further ascend Jacob's ladder of xc functionals~\cite{perdew2001jacob}, but in the ground state.

Moving on to the warm dense matter regime, temperature-effects in the xc functional become more important, and we find deviations in the electronic pressure clearly exceeding $5\%$. Moreover, these deviations are non monotonous with respect to $T$, and we find a sign change in the pressure difference, which is shifted to larger temperatures with increasing density. We thus conclude that the further development of xc functionals to consistently take into account thermal excitations is of central importance to achieve predictive capability for DFT calculations in the WDM regime.

In addition, we have presented the first finite-$T$ DFT results for hydrogen in the strongly coupled electron liquid regime, $r_s\gtrsim10$. At these conditions, electronic xc effects are even more important for an accurate description, and, consequently, the temperature-dependence of the xc functional is crucial. More specifically, we find pressure differences between the PZ and GDSMFB functionals exceeding $20\%$ at $r_s=14$ in the vicinity of the Fermi temperature. We expect this point to be of high importance for the future investigation of interesting phenomena such as the possible emergence of an incipient excitonic mode, which might occur at even lower density~\cite{PhysRevB.94.245106}.

Finally, we have extended our consideration to other physical properties of hydrogen like the density of states, and the electronic density in coordinate space.
Regarding the DOS, we have found that finite-$T$ xc effects do indeed significantly influence the DFT results in the WDM regime, as it is expected. Our simulation results for the electronic iso-surfaces in coordinate space are even more remarkable, as the PZ functional is not capable to describe the reduction of electronic overlap at finite temperature.

Possible topics for future research include the consideration of other materials such as helium or carbon (see Ref.~\cite{2019arXiv191209884B} for first results) and the investigation of transport properties like the electrical conductivity.

As a concluding remark, we note that thermal xc effects are highly important for many applications other than DFT, such as quantum hydrodynamics~\cite{zhandos_pop18,bonitz_pop_19} and astrophysical models~\cite{reefId0,saumon1995equation,Becker_2014}.


\begin{acknowledgments}
 We are grateful to M. Bonitz for helpful comments and M.A.L. Marques for implementing the GDSMFB xc functional in LIBXC. KR would like to thank M. Bonitz for the hospitality at ITAP Kiel and also thank S. Groth for the stimulating discussions. TD acknowledges funding by the Center of Advanced Systems Understanding (CASUS) which is financed by Germany’s Federal Ministry of Education and Research (BMBF) and by the Saxon Ministry for Science, Culture and Tourism (SMWK) with tax funds on the basis of the budget approved by the Saxon State Parliament.
Computations were performed on a Bull Cluster at the Center for Information Services and High Performance Computing (ZIH) at TU Dresden. We would like to thank the ZIH for its support and generous allocations of computer time.
\end{acknowledgments}   

\section*{Appendix}

\appendix

\section{Phase diagram with LLPT boundaries\label{sec:appendix_phase}}

\begin{figure}[th]     
\centering 
\includegraphics[width=1.0\columnwidth]{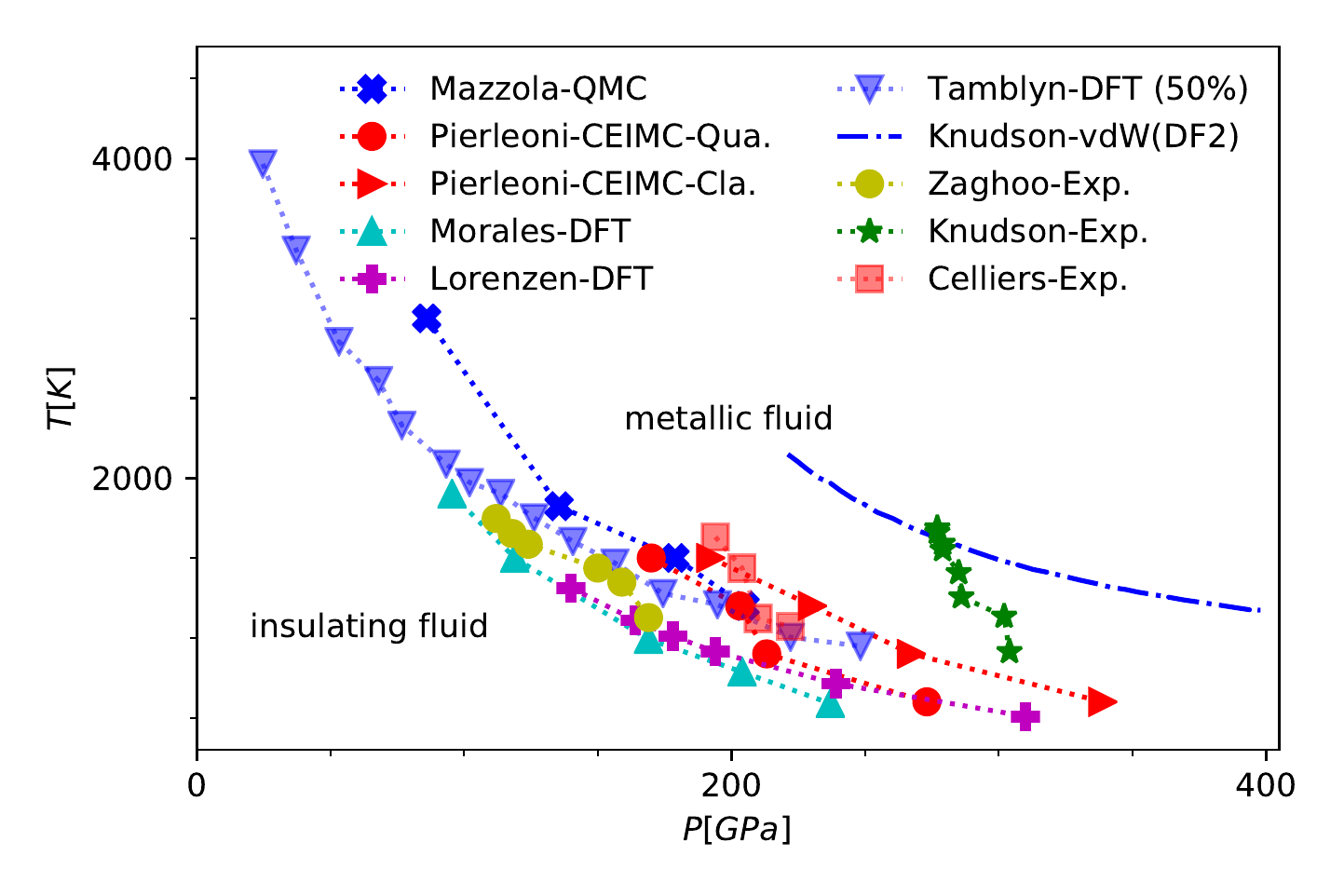}  
\caption{\raggedright The phase diagram of hydrogen at high densities. Mazzola Ref.~\onlinecite{PhysRevLett.120.025701}; Morales Ref.~\onlinecite{PhysRevE.81.021202}; Pierleoni Ref.~\onlinecite{Pierleoni4953, pierleoni2018local}; Lorenzen Ref.~\onlinecite{PhysRevB.82.195107}; Tamblyn Ref.~\onlinecite{PhysRevLett.104.065702};  Zaghoo Ref.~\onlinecite{PhysRevB.93.155128}; Knudson Ref.~\onlinecite{Knudson1455}; Celliers Ref.~\onlinecite{Celliers677} }  
\label{H2_phase_diagram}          
\end{figure}     

In Fig.~\ref{H2_phase_diagram}, we show the phase diagram of hydrogen at high densities including the LLPT. The LDA (PZ/GDSMFB) results of this work are not included into the diagram as they fail to capture the LLPT and the EOS is inconsistent compared to other xc functionals and experimental results ~\cite{PhysRevLett.120.025701, PhysRevE.81.021202, Pierleoni4953, pierleoni2018local, PhysRevB.82.195107, PhysRevLett.104.065702, PhysRevB.93.155128, Knudson1455, Celliers677}. The Pierleoni \textit{et al.} data are shown both for quantum and classical protons~\cite{Pierleoni4953,pierleoni2018local}.  


\section{Convergence in high density limit} 
\label{app_low_rs}  

\begin{figure}[H]       
\centering 
\includegraphics[width=1.0\columnwidth]{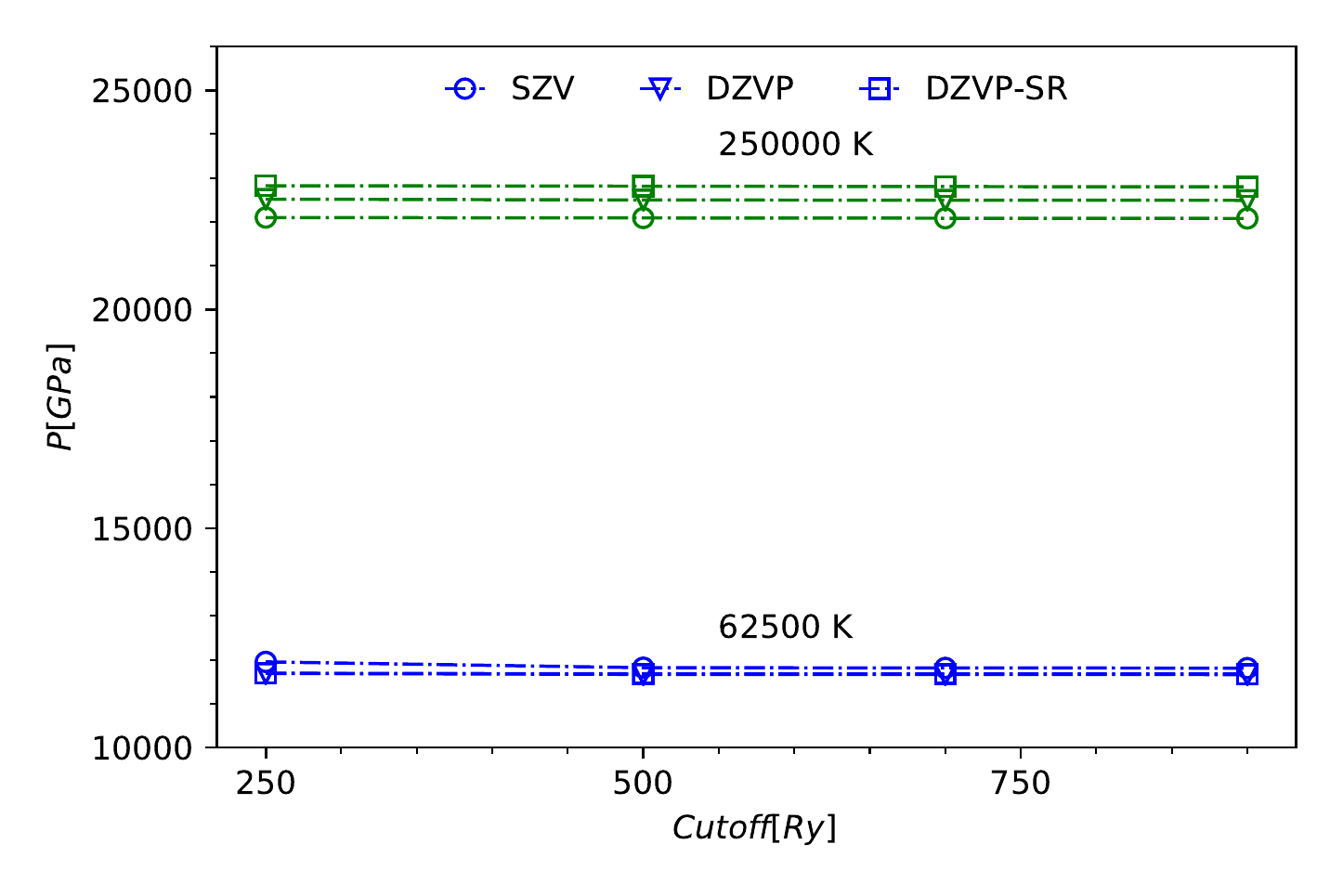}  
\caption{\raggedright The convergence of pressure with respect to the plane wave energy cutoff and basis sets at 62500 K and 250000 K for $r_{s}=0.8137$. }   
\label{H2_rs_0.7_conv}            
\end{figure}

In Fig. \ref{H2_rs_0.7_conv}, the energy cutoff for $r_{s}=0.8137$ using various basis sets with PZ exchange correlation for 32 atoms is shown at two different temperatures. A cutoff of 500 Ry and above ensures the convergence in pressure calculations.  
  
\section{Convergence in low density limit} 
\label{app_high_rs} 
\begin{figure}[H]     
\centering 
\includegraphics[width=1.0\columnwidth]{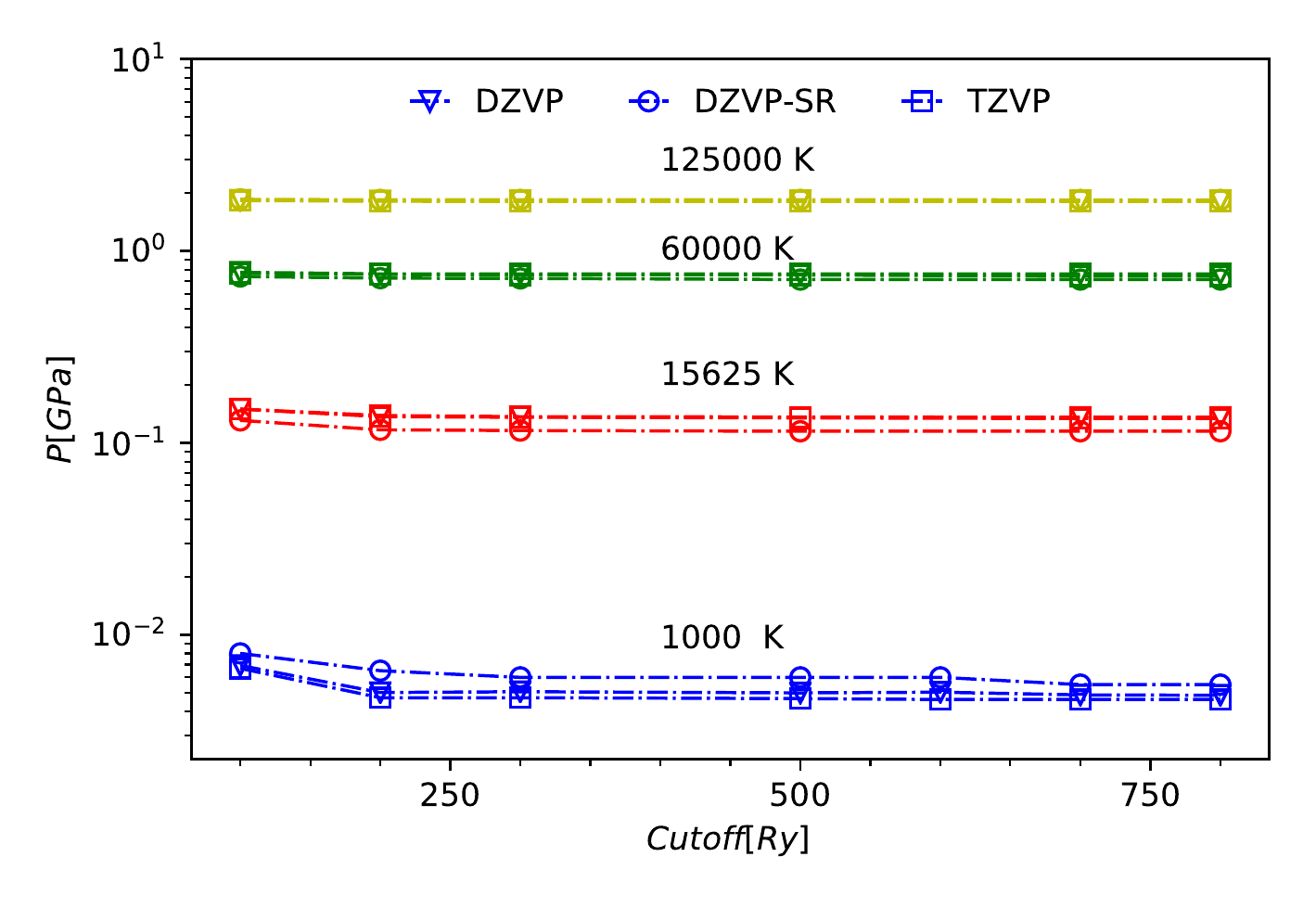}  
\caption{\raggedright The convergence of pressure with respect to the plane wave energy cutoff and basis sets at various temperatures for $r_{s}=14$. }   
\label{H2_rs_14_conv}           
\end{figure}

In Fig. \ref{H2_rs_14_conv}, the energy cutoff for $r_{s}=14$ using various basis sets with PZ exchange correlation for 32 atoms is shown at four different temperatures. A cutoff of 500 Ry and above ensures the convergence in pressure calculations.  

\bibliography{Metallic-Hydrogen-bibliography}

\begin{thebibliography}{119}
\bibitem [{mis()}]{misc}%
  \BibitemOpen
  \href@noop {} {}\bibinfo {note} {We note that this parametrization is
  sometimes being referred to by "GDB". Here, we choose to follow the more
  common custom and include all authors into the abbreviation.}\BibitemShut
  {Stop}%
\end{thebibliography}

\end{document}